\documentclass{llncs}

\usepackage{graphicx}
\usepackage[tight,footnotesize]{subfigure}
\usepackage{url}
\usepackage{tikz} 
\usepackage{algorithm}     % algorithms
\usepackage{algorithmic}     % more algorithms
\usepackage{graphicx}     % advanced figures
\usepackage{xspace}     % fix space in macros
\usepackage{booktabs}     % nicer tables
\usepackage[bf,tableposition=top]{caption}     % captions on top for tables
\usepackage[square,numbers]{natbib}     % better references
\usepackage{siunitx}          % \num for decimal grouping

\usepackage{multirow}
\usepackage{subfloat}

\usepackage{listings}
\usepackage{color}

\usepackage[T1]{fontenc}
\usepackage{lmodern}
\usepackage{colortbl}
\definecolor{myblue}{rgb}{0.8,0.85,1}

\long\def\BEGINOMIT#1\ENDOMIT{\relax}  % to omit large portions of text

\newcommand{\samoa}{Apache \textsc{samoa}\xspace}
\newcommand{\moa}{\textsc{moa}\xspace}

\newcommand{\vht}{VHT\xspace}

\newcommand{\vfdt}{VFDT\xspace}

\newcommand{\wok}{\textbf{wok}\xspace}
\newcommand{\wk}[1]{\textbf{wk(#1)}\xspace}
\newcommand{\wkz}{\wk{z}}

\newcommand{\spara}[1]{\smallskip\noindent\textbf{#1}}

\newenvironment {squishlist}
{\begin{list}{$\bullet$}
  { \setlength{\itemsep}{0pt}
     \setlength{\parsep}{3pt}
     \setlength{\topsep}{3pt}
     \setlength{\partopsep}{0pt}
     \setlength{\leftmargin}{1.5em}
     \setlength{\labelwidth}{1em}
     \setlength{\labelsep}{0.5em} } }
{\end{list}}

\begin{document}

\title{Large-Scale Learning from Data Streams with Apache SAMOA}

\titlerunning{Stream Mining with Apache Samoa}  % abbreviated title (for running head)
\author{Nicolas Kourtellis\inst{1} \and Gianmarco De Francisci Morales\inst{2}
\and Albert Bifet\inst{3}}
\authorrunning{Nicolas Kourtellis et al.} % abbreviated author list (for running head)

\institute{
Telefonica Research, Spain, 
\email{nicolas.kourtellis@telefonica.com}
\and 
Qatar Computing Research Institute, Qatar, 
\email{gdfm@acm.org}
\and 
LTCI, T\'el\'ecom ParisTech, France, 
\email{albert.bifet@telecom-paristech.fr}
}

\maketitle
\begin{abstract}
Apache SAMOA (Scalable Advanced Massive Online Analysis) is an open-source platform for mining big data streams. 
Big data is defined as datasets whose size is beyond the ability of typical software tools to capture, store, manage, and analyze, due to the time and memory complexity.
Apache SAMOA provides a collection of distributed streaming algorithms for the most common data mining and machine learning tasks such as classification, clustering, and regression, as well as programming abstractions to develop new algorithms.
It features a pluggable architecture that allows it to run on several distributed stream processing engines such as Apache Flink, Apache Storm, and Apache Samza.
Apache SAMOA is written in Java and is available at \url{https://samoa.incubator.apache.org} under the Apache Software License version 2.0.
\end{abstract}

\section{Introduction}

Big data are ``data whose characteristics force us to look beyond the traditional methods that are prevalent at the time''~\cite{jacobs2009pathologies}.
For instance, social media are one of the largest and most dynamic sources of data.
These data are not only very large due to their fine grain, but also being produced continuously.
Furthermore, such data are nowadays produced by users in different environments and via a multitude of devices.
For these reasons, data from social media and ubiquitous environments are perfect examples of the challenges posed by big data.

Currently, there are two main ways to deal with these challenges: streaming algorithms and distributed computing (e.g., MapReduce).
\samoa aims at satisfying the future needs for big data stream mining by combining the two approaches in a single platform under an open source umbrella~\cite{deFrancisciMorales2014samoa}.

Data mining and machine learning are well established techniques among Web and social media companies to draw insights from data coming from ubiquitous and social environments.
Online content analysis for detecting aggression~\cite{chatzakou2017websci-bullying}, stock trade volume prediction~\cite{bordino2014stocks}, online spam detection~\cite{chen2015spam-groundtruth-tweets},
recommendation~\cite{devooght2015matrix}, and personalization~\citep{de2012chatter} are just a few of the applications made possible by mining the huge quantity of data available nowadays.
Just think of Facebook's relevance algorithm for the news feed for a famous example.

In order to cope with Web-scale datasets, data scientists have resorted to \emph{parallel and distributed computing}.
MapReduce~\cite{dean2004mapreduce} is currently the de-facto standard programming paradigm in this area,
mostly thanks to the popularity of Hadoop\footnote{\url{http://hadoop.apache.org}}, an open source implementation of MapReduce.
Hadoop and its ecosystem (e.g., Mahout\footnote{\url{http://mahout.apache.org}}) have proven to be an extremely successful platform to support the aforementioned process at web scale.

However, nowadays most data are generated in the form of a stream, especially when dealing with social media.
Batch data are just a snapshot of streaming data obtained in an interval (window) of time.
Researchers have conceptualized and abstracted this setting in the \emph{streaming model}.
In this model, data arrive at high speed, one instance at a time, and algorithms must process it in one pass under very strict constraints of space and time.
Extracting knowledge from these massive data streams to perform dynamic network analysis~\cite{kourtellis15scalable-online-betweenness} or to create predictive models~\cite{aggarwal2007datastreams}, and using them, e.g., to choose a suitable business strategy, or to improve healthcare services, can generate substantial competitive advantages.
Many applications need to process incoming data and react on-the-fly by using comprehensible prediction mechanisms (e.g., card fraud detection) and, thus, streaming algorithms make use of probabilistic data structures to give fast and approximate answers.

On the one hand, MapReduce is not suitable to express streaming algorithms.
On the other hand, traditional sequential online algorithms are limited by the memory and bandwidth of a single machine.
\emph{Distributed stream processing engines} (DSPEs) are a new emergent family of MapReduce-inspired technologies that address this issue.
These engines allow to express parallel computation on streams, and combine the scalability of distributed processing with the efficiency of streaming algorithms.
Examples include Storm\footnote{\url{http://storm.apache.org}}, Flink\footnote{\url{http://flink.apache.org}},  Samza\footnote{\url{http://samza.apache.org}}, and Apex\footnote{\url{https://apex.apache.org}}.

\begin{figure}[t]
\centering
\includegraphics[width=7cm]{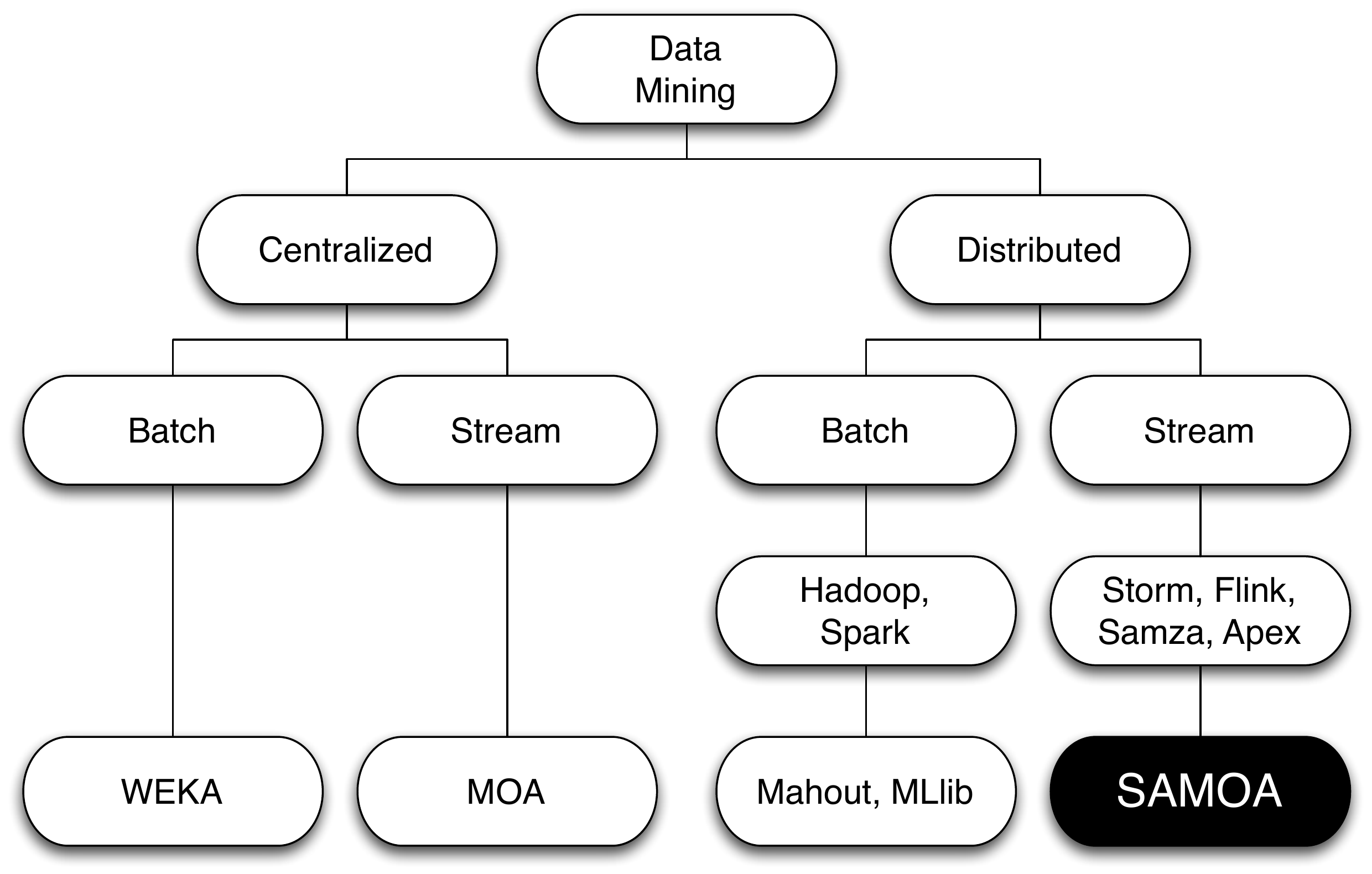}
\includegraphics[width=5cm]{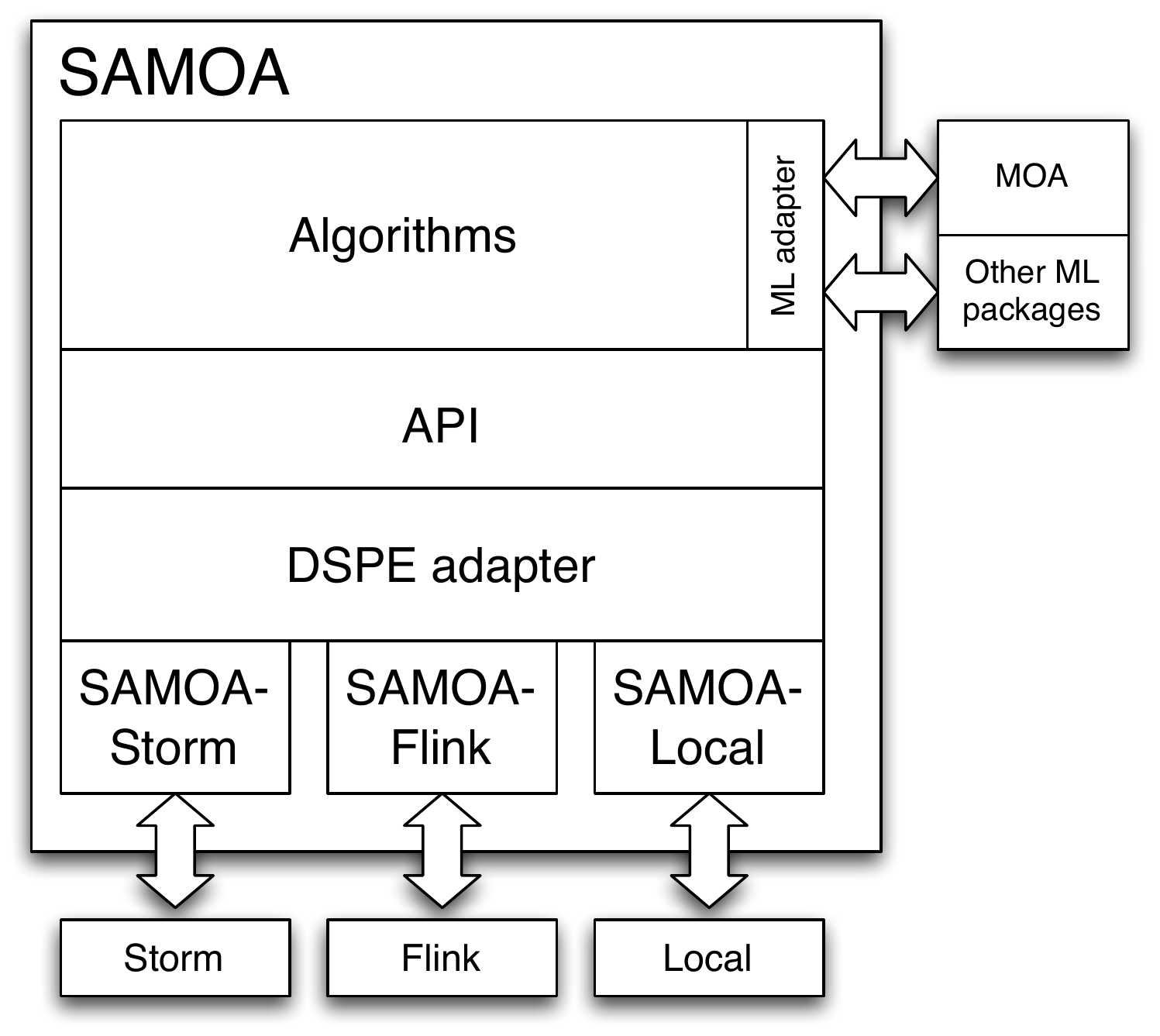}
\caption{(Left) Taxonomy of data mining and machine learning tools. (Right) High level architecture of \samoa.}
\label{fig:taxonomy_samoa_architecture}
\vspace{-5mm}
\end{figure}

Alas, currently there is no common solution for mining big data streams, that is, for running data mining and machine learning algorithms on a distributed stream processing engine.
The goal of \samoa is to fill this gap, as exemplified by Figure~\ref{fig:taxonomy_samoa_architecture}(left).

\section{Description}

\samoa \textsc{(Scalable Advanced Massive Online Analysis)} is a platform for mining big data streams~\cite{deFrancisciMorales2013samoa}.
For a simple analogy, think of \samoa as Mahout for streaming.
As most of the rest of the big data ecosystem, it is written in Java.

\samoa is both a framework and a library.
As a framework, it allows the algorithm developer to abstract from the underlying execution engine, and therefore reuse their code on different engines.
It features a pluggable architecture that allows it to run on several distributed stream processing engines such as Storm, Flink, Samza, and Apex.
This capability is achieved by designing a minimal API that captures the essence of modern DSPEs.
This API also allows to easily write new bindings to port \samoa to new execution engines.
\samoa takes care of hiding the differences of the underlying DSPEs in terms of API and deployment.

As a library, \samoa contains implementations of state-of-the-art algorithms for distributed machine learning on streams.
For classification, \samoa provides a Vertical Hoeffding Tree (VHT), a distributed streaming version of a decision tree.
For clustering, it includes an algorithm based on CluStream.
For regression, HAMR, a distributed implementation of Adaptive Model Rules.
The library also includes meta-algorithms such as bagging and boosting~\citep{vasiloudis2017boostvht}.
The platform is intended to be useful for both research-oriented settings for the design and experimentation of new algorithms, and real world deployments in production settings.

\spara{Related Work.}
\label{sec:rel-work}
We identify two frameworks that belong to the category of distributed streaming machine learning: Jubatus and StormMOA.
Jubatus\footnote{\url{http://jubat.us/en}} is an example of distributed streaming machine learning framework.
It includes a library for streaming machine learning such as regression, classification, recommendation, anomaly detection and graph mining.
It introduces the local ML model concept which means there can be multiple models running at the same time and they process different sets of data.
Using this technique, Jubatus achieves scalability via horizontal parallelism in partitioning data.
We test horizontal parallelism in our experiments, by implementing a horizontally scaled version of the hoeffding tree.
Jubatus establishes tight coupling between the machine learning library implementation and the underlying distributed stream processing engine (SPE).
The reason is Jubatus builds and implements its own custom distributed SPE.
In addition, Jubatus does not offer any tree learning algorithm, as all of its models need to be linear by construction.

StormMOA\footnote{\url{http://github.com/vpa1977/stormmoa}} is a project to combine MOA with Storm to satisfy the need of scalable implementation of streaming ML frameworks.
It uses Storm's Trident abstraction and MOA library to implement OzaBag and OzaBoost\cite{oza_online_2001}.
Similarly to Jubatus, StormMOA also establishes tight coupling between MOA (the machine learning library) and Storm (the underlying distributed SPE).
This coupling prevents StormMOA's extension by using other SPEs to execute the machine learning library.
StormMOA only allows to run a single model in each Storm bolt (processor).
This characteristic restricts the kind of models that can be run in parallel to ensembles.

\section{High Level Architecture}
\label{sec:high_level_architecture}

We identify three types of \samoa users:
\begin{enumerate}
\item Platform users, who use available ML algorithms without implementing new ones. 
\item ML developers, who develop new ML algorithms on top of \samoa and want to be isolated from changes in the underlying SPEs. 
\item Platform developers, who extend \samoa to integrate more DSPEs into \samoa.
\end{enumerate}

There are three important design goals of \samoa:
\begin{enumerate}
\item \textbf{Flexibility} in term of developing new ML algorithms or reusing existing ML algorithms from other frameworks.
\item \textbf{Extensibility} in term of porting \samoa to new DSPEs.
\item \textbf{Scalability} in term of handling ever increasing amount of data.
\end{enumerate}

Figure~\ref{fig:taxonomy_samoa_architecture}(right) shows the high-level architecture of \samoa which attempts to fulfill the aforementioned design goals.
The \textit{algorithm} layer contains existing distributed streaming algorithms that have been implemented in \samoa.
This layer enables platform users to easily use the existing algorithm on any DSPE of their choice.

The \textit{application programming interface} (API) layer consists of primitives and components that facilitate ML developers when implementing new algorithms. The \textit{ML-adapter} layer allows ML developers to integrate existing algorithms in \moa or other ML frameworks into \samoa. The API layer and ML-adapter layer in \samoa fulfill the flexibility goal since they allow ML developers to rapidly develop algorithms.

Next, the \textit{DSPE-adapter} layer supports platform developers in integrating new DSPEs into \samoa.
To perform the integration, platform developers should implement the \textit{samoa-SPE} layer as shown in Figure~\ref{fig:taxonomy_samoa_architecture}(right).
Currently \samoa is equipped with four adapters: the \textit{samoa-Storm} adapter for Storm, the \textit{samoa-Samza} adapter for Samza, the \textit{samoa-Flink} adapter for Flink, and the \textit{samoa-Apex} adapter for Apex.
To satisfy the extensibility goal, the DSPE-adapter layer decouples DSPEs and ML algorithms implementations in \samoa, so that platform developers are able to easily integrate more DSPE platforms.

The last goal, scalability, implies that \samoa should be able to scale to cope ever increasing amount of data. To fulfill this goal, \samoa utilizes modern DSPEs to execute its ML algorithms.
The reason for using modern DSPEs such as Storm, Flink, Samza and Apex in \samoa is that they are designed to provide horizontal scalability to cope with Web-scale streams.

\section{System Design}

An algorithm in \samoa is represented by a directed graph of nodes that communicate via messages along streams which connect pairs of nodes.
Borrowing the terminology from Storm, this graph is called a \emph{Topology}.
Each node in a Topology is a \emph{Processor} that sends messages through a \emph{Stream}.
A Processor is a container for the code implementing the algorithm.
A Stream can have a single source but several destinations (akin to a pub-sub system).
A Topology is built by using a \emph{TopologyBuilder}, which connects the various pieces of user code to the platform code and performs the necessary bookkeeping in the background.
The following code snippet builds a topology that joins two data streams in \samoa:

\begin{lstlisting}[ keywordstyle=\ttfamily\color{blue}, morekeywords={TopologyBuilder, Processor, Stream}]
TopologyBuilder builder = new TopologyBuilder();
Processor sourceOne = new SourceProcessor();
builder.addProcessor(sourceOne);
Stream streamOne = builder
	.createStream(sourceOne);

Processor sourceTwo = new SourceProcessor();
builder.addProcessor(sourceTwo);
Stream streamTwo = builder
	.createStream(sourceTwo);

Processor join = new JoinProcessor();
builder.addProcessor(join)
	.connectInputShuffle(streamOne)
	.connectInputKey(streamTwo);
\end{lstlisting}

A \emph{Task} is an execution entity, similar to a job in Hadoop.
A Topology is instantiated inside a Task to be run by \samoa.
An example of a Task is \emph{PrequentialEvaluation}, a classification task where each instance is used for testing first, and then for training.

A message or an event is called \emph{Content Event} in \samoa.
As the name suggests, it is an event which contains content that needs to be processed by the processors.
Finally, a \emph{Processing Item} is a hidden physical unit of the topology and is just a wrapper of Processor. It is used internally, and it is not accessible from the API.

\section{Machine Learning Algorithms}

In \samoa there are currently three types of algorithms performing basic machine learning functionalities such as classification via a decision tree (VHT), clustering (CluStream) and regression rules (AMR).

The Vertical Hoeffding Tree (VHT)~\cite{kourtellis2016vht} is a distributed extension of the VFDT~\cite{Domingos2000vfdt}.
VHT uses vertical parallelism to split the workload across several machines.
Vertical parallelism leverages the parallelism across attributes in the same example, rather than across different examples in the stream.
In practice, each training example is routed through the tree model to a leaf.
There, the example is split into its constituting attributes, and each attribute is sent to a different Processor instance that keeps track of sufficient statistics.
This architecture has two main advantages over one based on horizontal parallelism.
First, attribute counters are not replicated across several machines, thus, reducing the memory footprint.
Second, the computation of the fitness of an attribute for a split decision (via, e.g., entropy or information gain) can be performed in parallel.
The drawback is that in order to get good performances, there must be sufficient inherent parallelism in the data.
That is, the VHT works best for high-dimensional data.

\samoa includes a distributed version of CluStream, an algorithm for clustering evolving data streams.
CluStream keeps a small summary of the data received so far by computing micro-clusters online.
These micro-clusters are further refined to create macro-clusters by a micro-batch process, which is triggered periodically.
The period is configured via a command line parameter (e.g., every $10\,000$ examples).

For regression, \samoa provides a distributed implementation of Adaptive Model Rules~\cite{thuvu2014amrules}.
The algorithm, HAMR, uses a hybrid of vertical and horizontal parallelism to distribute AMRules on a cluster.

\samoa also includes adaptive implementations of ensemble methods such as bagging and boosting.
These methods include state-of-the-art change detectors such as as ADWIN, DDM, EDDM, and Page-Hinckley~\cite{survey}.
These meta-algorithms are most useful in conjunction with external single-machine classifiers, which can be plugged in \samoa in several ways.
For instance, open-source connectors for \moa~\cite{bifet2010moa} are provided separately by the \samoa-\moa package.\footnote{\url{https://github.com/samoa-moa/samoa-moa}}

The following listing shows how to download, build, and run \samoa.
\begin{lstlisting}[language=bash,morekeywords={mvn,git}]
# download and build SAMOA
git clone http://git.apache.org/incubator-samoa.git
cd incubator-samoa
mvn package    

# download the Forest Cover Type dataset
wget "http://downloads.sourceforge.net/project/moa-datastream/Datasets/Classification/covtypeNorm.arff.zip"
unzip "covtypeNorm.arff.zip"

# run SAMOA in local mode
bin/samoa local target/SAMOA-Local-0.4.0-SNAPSHOT.jar "PrequentialEvaluation -l classifiers.ensemble.Bagging -s (ArffFileStream -f covtypeNorm.arff) -f 100000"
\end{lstlisting}

\section{Vertical Hoeffding Tree}
\label{sec:algorithm}

We explain the details of the \emph{Vertical Hoeffding Tree}~\cite{kourtellis2016vht}, which is a data-parallel, distributed version of the Hoeffding tree.
The \emph{Hoeffding tree}~\cite{Domingos2000vfdt} (a.k.a. \vfdt) is a streaming decision tree learner with statistical guarantees.
In particular, by leveraging the Chernoff-Hoeffding bound~\cite{Hoeffding1963bound}, it guarantees that the learned model is asymptotically close to the model learned by the batch greedy heuristic, under mild assumptions.
The learning algorithm is very simple.
Each leaf keeps track of the statistics for the portion of the stream it is reached by, and computes the best two attributes according to the splitting criterion.
Let $\Delta G$ be the difference between the value of the functions that represent the splitting criterion of these two attributes.
Let $\epsilon$ be a quantity that depends on a user-defined confidence parameter $\delta$, and that decreases with the number of instances processed.
When $\Delta G > \epsilon$, then the current best attribute is selected to split the leaf.
The Hoeffding bound guarantees that this choice is the correct one with probability at least $1 - \delta$.

In this section, first, we describe the parallelization and the ideas behind our design choice.
Then, we present the engineering details and optimizations we employed to obtain the best performance.

\subsection{Vertical Parallelism}
\label{sec:vertical_parallelism}

Data parallelism is a way of distributing work across different nodes in a parallel computing environment such as a cluster.
In this setting, each node executes the same operation on different parts of the dataset.
Contrast this definition with task parallelism (aka pipelined parallelism), where each node executes a different operator, and the whole dataset flows through each node at different stages.
When applicable, data parallelism is able to scale to much larger deployments, for two reasons:
($i$) data has usually much higher intrinsic parallelism that can be leveraged compared to tasks,
and ($ii$) it is easier to balance the load of a data-parallel application compared to a task-parallel one.
These attributes have led to the high popularity of the currently available DSPEs.
For these reasons, we employ data parallelism in the design of \vht.

\begin{figure}
	\centering
	\includegraphics[width=0.8\columnwidth]{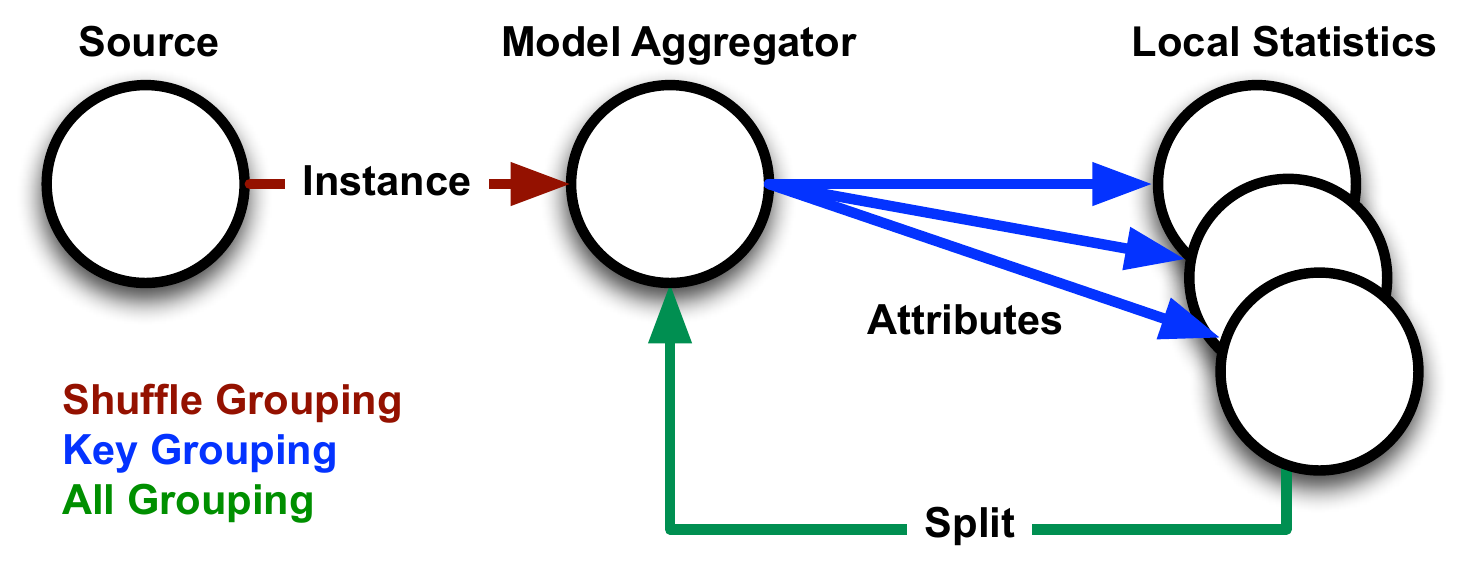}
	\caption{High level diagram of the \vht topology.}
	\label{fig:vht-topology}
\vspace{-4mm}
\end{figure}

In machine learning, it is common to think about data in matrix form.
A typical linear classification formulation requires to find a vector $x$ such that $A \cdot x \approx b$, where $A$ is the data matrix and $b$ is a class label vector.
The matrix $A$ is $n \times m$-dimensional, with $n$ being the number of data instances and $m$ being the number of attributes of the dataset.

There are two ways to \emph{slice} the data matrix to obtain data parallelism: by row or column.
The former is called \emph{horizontal parallelism}, the latter \emph{vertical parallelism}.
With horizontal parallelism, data instances are independent from each other, and can be processed in isolation while considering all available attributes.
With vertical parallelism, instead, attributes are considered independently from each other.

The fundamental operation of the algorithm is to accumulate statistics $n_{ijk}$ (i.e., counters) for triplets of attribute $i$, value $j$, and class $k$, for each leaf $l$ of the tree.
The counters for each leaf are independent, so let us consider the case for a single leaf.
These counters, together with the learned tree structure, constitute the state of the \vht algorithm.

Different kinds of parallelism distribute the counters across computing nodes in different ways.
With horizontal parallelism~\cite{ben-haim2010spdt}, the instances are distributed randomly, thus multiple instances of the same counter can exist on several nodes.
On the other hand, when using vertical parallelism, the counters for one attribute are grouped on a single node.

This latter design has several advantages.
First, by having a single copy of the counter, the memory requirements for the model are the same as in the sequential version.
In contrast, with horizontal parallelism a single attribute may be tracked on every node, thus the memory requirements grow linearly with the parallelism level.
Second, by having each attribute tracked independently, the computation of the split criterion can be performed in parallel by several nodes.
Conversely, with horizontal partitioning the algorithm needs to (centrally) aggregate the partial counters before being able to compute the splitting criterion.

Of course, the vertically-parallel design has also its drawbacks.
In particular, horizontal parallelism achieves a good load balance more easily, even though solutions for these problems have recently been proposed for vertical parallelism as well~\cite{Uddin2015pobc,Uddin2016twochoices}.
In addition, if the instance stream arrives in row-format, it needs to be transformed in column-format, and this transformation generates additional CPU overhead at the source.
Indeed, each attribute that constitutes an instance needs to be sent independently, and needs to carry the class label of its instance.
Therefore, both the number of messages and the size of the data transferred increase.
Nevertheless, the advantages of vertical parallelism outweigh its disadvantages for several real-world settings.

\subsection{Algorithm Structure}

\begin{table}[tbp]
\caption{Definitions and symbols.}
\vspace{-\baselineskip}
\begin{center}
\footnotesize
\tabcolsep=0.5em
\begin{tabular}{l l l l}
\toprule
Description						&	Symbol	\\
\midrule
Source receiving/distributing instances	&	$S$			\\
Training instance from $S$			&	$E$			\\
Model Aggregator					&	$MA$		\\
Current state of the decision tree in MA	&	$VHT\_tree$	\\
Local Statistic						&	$LS$			\\
Counter for attribute i, value j, class k	&	$n_{ijk}$		\\
Instances seen at leaf $l$			 	&	$n_l$		\\
Information gain or entropy of attribute $i$ in leaf $l$	&	$\overline{G}_{l}(X^{local}_{i})$	\\

\bottomrule
\end{tabular}
\end{center}
\label{tab:definitions}
\vspace{-1mm}
\end{table}%

We are now ready to explain the structure of the \vht algorithm.
In general, there are two main parts to the Hoeffding tree algorithm: \emph{sorting} the instances through the current model, and accumulating \emph{statistics} of the stream at each leaf node.
This separation offers a neat cut point to modularize the algorithm in two separate components.
We call the first component \emph{model aggregator}, and the second component \emph{local statistics}.
Figure~\ref{fig:vht-topology} presents a visual depiction of the algorithm, and specifically, of its components and how the data flow among them.
Also, Table~\ref{tab:definitions} summarizes a list of components used in the rest of the algorithm description.

The model aggregator holds the current model (the tree) produced so far in a Processor node.
Its main duty is to receive the incoming instances and sort them to the correct leaf.
If the instance is unlabeled, the model predicts the label at the leaf and sends it downstream (e.g., for evaluation).
Otherwise, if the instance is labeled it is also used as training data.
The \vht decomposes the instance into its constituent attributes, attaches the class label to each, and sends them independently to the following stage, the \emph{local statistics}.
Algorithm~\ref{alg:vht_tree_induction} shows a pseudocode for the model aggregator.

\begin{algorithm}[t]
\caption{$MA$: Vertical\-Hoeffding\-TreeInduction($E$, $VHT\_tree$)}
\begin{algorithmic}[1]
{\footnotesize \REQUIRE $E$ wrapped in \verb;instance; content event
\REQUIRE $VHT\_tree$ in $MA$
\STATE Use $VHT\_tree$ to sort $E$ into a leaf $l$
\STATE Send \verb;attribute; content events to $LS$s
\STATE Increment $n_l$
\IF{$n_l$ $mod$ $n_{min}$ $= 0$ {\bf and} not all instances seen at $l$ belong to the same class}
\STATE Add $l$ into the list of splitting leaves
\STATE Send \verb;compute; content event with the id of leaf $l$ to all $LS$s
\ENDIF
}\end{algorithmic}
\label{alg:vht_tree_induction}
\end{algorithm}

The local statistics contain the counters $n_{ijk}$ for a set of attribute-value-class triplets.
Conceptually, the local statistics can be viewed as a large distributed table, indexed by leaf id (row), and attribute id (column).
The value of the cell represents a set of counters, one for each pair of attribute value and class.
The local statistics accumulate statistics on the data sent by the model aggregator.
Pseudocode for the update function is shown in Algorithm~\ref{alg:update_local_statistic}.

\begin{algorithm}[t]
\caption{$LS$: Update\-Local\-Statistic($attribute$, $local\_statistic$)}
\begin{algorithmic}[1]
{\footnotesize \REQUIRE $attribute$ is an \verb;attribute; content event
\REQUIRE $local\_statistic$ is the $LS$ in charge of $attribute$, could be implemented as\\ $Table<leaf\_id, attribute\_id>$
\STATE Update $local\_statistic$ with data in $attribute$: attribute value, class value and instance weights
}\end{algorithmic}
\label{alg:update_local_statistic}
\end{algorithm}

In \samoa, we implement vertical parallelism by connecting the model to the statistics via key grouping.
We use a composite key made by the leaf id and the attribute id.
Horizontal parallelism can similarly be implemented via shuffle grouping on the instances themselves.

\spara{Leaf splitting.}
Periodically, the model aggregator tries to see if the model needs to evolve by splitting a leaf.
When a sufficient number of instances $n_{min}$ have been sorted through a leaf, and not all instances that reached $l$ belong to the same class (line 4, Algorithm~\ref{alg:vht_tree_induction}), the aggregator sends a broadcast message to the statistics, asking to compute the split criterion for the given leaf id.
The statistics Processor gets the table corresponding to the leaf, and for each attribute compute the splitting criterion in parallel (an information-theoretic function such as information gain or entropy).
Each local statistic then sends back to the model the top two attributes according to the chosen criterion, together with their scores ($\overline{G}_{l}(X^{local}_{i}), i = a, b$; Algorithm~\ref{alg:receive_compute_message}).

\begin{algorithm}[t]
\caption{$LS$: Receive\-Compute\-Message($compute$, $local\_statistic$)}
\begin{algorithmic}[1]
{\footnotesize \REQUIRE $compute$ is a \verb;compute; content event
\REQUIRE $local\_statistic$ is the $LS$ in charge of $attribute$, could be implemented as\\ $Table<leaf\_id, attribute\_id>$
\STATE Get leaf $l$ ID from \verb;compute; content event
\STATE For each attribute $i$ that belongs to leaf $l$ in local statistic, compute $\overline{G}_{l}(X_{i})$
\STATE Find $X^{local}_{a}$, i.e., the attribute with highest $\overline{G}_{l}$ based on the local statistic
\STATE Find $X^{local}_{b}$, i.e., the attribute with second highest $\overline{G}_{l}$ based on the local statistic
\STATE Send $X^{local}_{a}$ and $X^{local}_{b}$ using \verb;local-result; content event to model-aggregator PI via \verb;computation-result; stream}
\end{algorithmic}
\label{alg:receive_compute_message}
\end{algorithm}

Subsequently, the model aggregator (Algorithm~\ref{alg:receive_local_result}) simply needs to compute the overall top two attributes received so far from the available statistics, apply the Hoeffding bound (line 4), and see whether the leaf needs to be split (line 5).
The algorithm also computes the criterion for the scenario where no split takes places ($X_{\emptyset}$).
\citet{Domingos2000vfdt} refer to this inclusion of a no-split scenario with the term \emph{pre-pruning}.
The decision to split or not is taken after a time has elapsed, as explained next.

By using the top two attributes, the model aggregator computes the difference of their splitting criterion values $\Delta \overline{G}_{l} = \overline{G}_{l}(X_{a}) - \overline{G}_{l}(X_{b})$.
To determine whether the leaf needs to be split, it compares the difference $\Delta \overline{G}_{l}$ to the Hoeffding bound $\epsilon = \sqrt{\frac{R^{2}\ln(1/\delta)}{2n_{l}}}$ for the current confidence parameter $\delta$ (where $R$ is the range of possible values of the criterion).
If the difference is larger than the bound ($\Delta \overline{G}_{l} > \epsilon$), then $X_{a}$ is the best attribute with high confidence $1-\delta$, and can therefore be used to split the leaf.
If the best attribute is the no-split scenario ($X_{\emptyset}$), the algorithm does not perform any split.
The algorithm also uses a tie-breaking $\tau$ mechanism to handle the case where the difference in splitting criterion between $X_{a}$ and $X_{b}$ is very small.
If the Hoeffding bound becomes smaller than $\tau$ ($\Delta \overline{G}_{l} <  \epsilon < \tau$), then the current best attribute is chosen regardless of the values of $\Delta \overline{G}_{l}$.

Two cases can arise: the leaf needs splitting, or it does not.
In the latter case, the algorithm simply continues without taking any action.
In the former case instead, the model modifies the tree by splitting the leaf $l$ on the selected attribute, replacing $l$ with an internal node (line 6), and generating a new leaf for each possible value of the branch (these leaves are initialized using the class distribution observed at the best attribute splitting at $l$ (line 8)).
Then, it broadcasts a \texttt{drop} message containing the former leaf id to the local statistics (line 10).
This message is needed to release the resources held by the leaf and make space for the newly created leaves.
Subsequently, the tree can resume sorting instances to the new leaves.
The local statistics creates a new table for the new leaves lazily, whenever they first receive a previously unseen leaf id.
In its simplest version, while the tree adjustment is performed, the algorithm drops the new incoming instances.
We show in the next section an optimized version that buffers them to improve accuracy.

\begin{algorithm}[t]
\caption{$MA$: Receive($local\_result$, $VHT\_tree$)}
\begin{algorithmic}[1]
{\footnotesize \REQUIRE $local\_result$ is an \verb;local-result; content event
\REQUIRE $VHT\_tree$ in $MA$
\STATE Get correct leaf $l$ from the list of splitting leaves
\STATE Update $X_{a}$ and $X_{b}$ in the splitting leaf $l$ with $X^{local}_{a}$ and $X^{local}_{b}$ from $local\_result$
\IF{$local\_results$ from all $LS$s received or time out reached}
\STATE Compute Hoeffding bound $\epsilon = \sqrt{\frac{R^{2}\ln(1/\delta)}{2n_{l}}}$
\IF{$X_{a} \ne X_{\emptyset}$ {\bf and} ($\overline{G}_{l}(X_{a}) - \overline{G}_{l}(X_{b}) > \epsilon$ {\bf or} $\epsilon < \tau$)}
\STATE Replace $l$ with a split-node on $X_{a}$
\FORALL{branches of the split}
\STATE Add new leaf with derived sufficient statistic from split node
\ENDFOR
\STATE Send \verb;drop; content event with id of leaf $l$ to all $LS$s
\ENDIF
\ENDIF}
\end{algorithmic}
\label{alg:receive_local_result}
\end{algorithm}

\spara{Messages.}
During the \vht execution several types of events are sent and received from the different parts of the algorithm, as summarized in Table~\ref{tab:content-events}.

\begin{table}[htbp]
\caption{Types of content events used during the execution of \vht algorithm.}
\vspace{-\baselineskip}
\begin{center}
\footnotesize
\tabcolsep=0.5em
\begin{tabular}{l l l l}
\toprule
Name			&	Parameters								&	From		&	To							\\
\midrule
\verb;instance;		&	$<$ attr 1, \dots, attr m, class $C >$				&	S		&	MA							\\
\verb;attribute; 		&	$<$ attr $id$, attr value, class $C >$				&	MA		&	LS $id = <$ leaf $id$ + attr $id >$	\\
\verb;compute;		&	$<$ leaf $id >$								&	MA 		&	All LS						\\
\verb;local-result;	&	$<$ $\overline{G}_{l}(X^{local}_{a}), \overline{G}_{l}(X^{local}_{b})$ $>$	&	$LS_{id}$	&	MA 			\\
\verb;drop;		&	$<$ leaf $id >$								&	MA		&	All LS						\\
\bottomrule
\end{tabular}
\end{center}
\label{tab:content-events}
\vspace{-10mm}
\end{table}%

\subsection{Evaluation}
\label{sec:experiments}

In our experimental evaluation of the \vht method, we aim to study the following questions:
\begin{squishlist}
\item[\bf Q1:] How does a centralized \vht compare to a centralized hoeffding tree with respect to accuracy and throughput?
\item[\bf Q2:] How does the vertical parallelism used by \vht compare to the horizontal parallelism?
\item[\bf Q3:] What is the effect of number and density of attributes?
\item[\bf Q4:] How does discarding or buffering instances affect the performance of \vht?
\end{squishlist}

\spara{Experimental setup.}
In order to study these questions, we experiment with five datasets (two synthetic generators and three real datasets), five different versions of the hoeffding tree algorithm, and up to four levels of computing parallelism.
We measure classification accuracy during, and at the end of the execution, and throughput (number of classified instances per second).
We execute each experimental configuration ten times and report the average of these measures.

\spara{Synthetic datasets.}
We use synthetic data streams produced by two random generators: one for dense and one for sparse attributes.

\begin{squishlist}
\item \textbf{Dense attributes} are extracted from a random decision tree.
We test different number of attributes, and include both categorical and numerical types.
The label for each configuration is the number of categorical-numerical used (e.g, 100-100 means the configuration has 100 categorical and 100 numerical attributes).
We produce 10 differently seeded streams with 1M instances for each tree, with one of two balanced classes in each instance, and take measurements every 100k instances.

\item \textbf{Sparse attributes} are extracted from a random tweet generator.
We test different dimensionalities for the attribute space: 100, 1k, 10k.
These attributes represent the appearance of words from a predefined bag-of-words.
On average, the generator produces 15 words per tweet (size of a tweet is Gaussian), and uses a Zipf distribution with skew $z=1.5$ to select words from the bag.
We produce 10 differently seeded streams with 1M tweets in each stream.
Each tweet has a binary class chosen uniformly at random, which conditions the Zipf distribution used to generate the words.
\end{squishlist}

\spara{Real datasets.}
We also test \vht on three real data streams to assess its performance on benchmark data.\footnote{\url{http://moa.cms.waikato.ac.nz/datasets/},\\\url{http://osmot.cs.cornell.edu/kddcup/datasets.html}}
\begin{squishlist}
\item ($elec$) Electricity: 45312 instances, 8 numerical attributes, 2 classes.
\item ($phy$) Particle Physics: 50000 instances, 78 numerical attributes, 2 classes.
\item ($covtype$) CovertypeNorm: 581012 instances, 54 numerical attributes, 7 classes.
\end{squishlist}

\spara{Algorithms.}
We compare the following versions of the hoeffding tree algorithm.

\begin{squishlist}
	\item \textbf{\moa}: This is the standard Hoeffding tree in \moa.
	\item \textbf{local}: This algorithm executes \vht in a local, sequential execution engine. All split decisions are made in a sequential manner in the same process, with no communication and feedback delays between statistics and model.
	\item \textbf{\wok}: This algorithm discards instances that arrive during a split decision.
This version is the vanilla \vht.
	\item \textbf{\wk{z}}: This algorithm sends instances that arrive during a split decision downstream.
In also adds instances to a buffer of size $z$ until full.
If the split decision is taken, it replays the instances in the buffer through the new tree model.
Otherwise, it discards the buffer, as the instances have already been incorporated in the statistics downstream.
	\item \textbf{sharding}: Splits the incoming stream horizontally among an ensemble of Hoeffding trees.
The final prediction is computed by majority voting.
This method is an instance of horizontal parallelism applied to Hoeffding trees.
It creates an ensemble of hoeffding trees, but each tree is built with a subset of instances split horizontally, while using all available attributes.
\end{squishlist}

\spara{Experimental configuration.}
All experiments are performed on a Linux server with 24 cores (Intel Xeon X5650), clocked at 2.67GHz, L1d cache: 32kB, L1i cache: 32kB, L2 cache: 256kB, L3 cache: 12288kB, and 65GB of main memory.
On this server, we run a Storm cluster (v0.9.3) and zookeeper (v3.4.6).
We use \samoa v0.4.0 (development version) and \moa v2016.04 available from the respective project websites.

We use several parallelism levels in the range of $p=2,\dots,16$, depending on the experimental configuration.
For dense instances, we stop at $p=8$ due to memory constraints, while for sparse instances we scale up to $p=16$.
We disable model replication (i.e., use a single model aggregator), as in our setup the model is not the bottleneck.

\subsubsection{Accuracy and time of \vht local vs. \moa.}

\begin{figure}[t]
\centering
\includegraphics[scale=1.05]{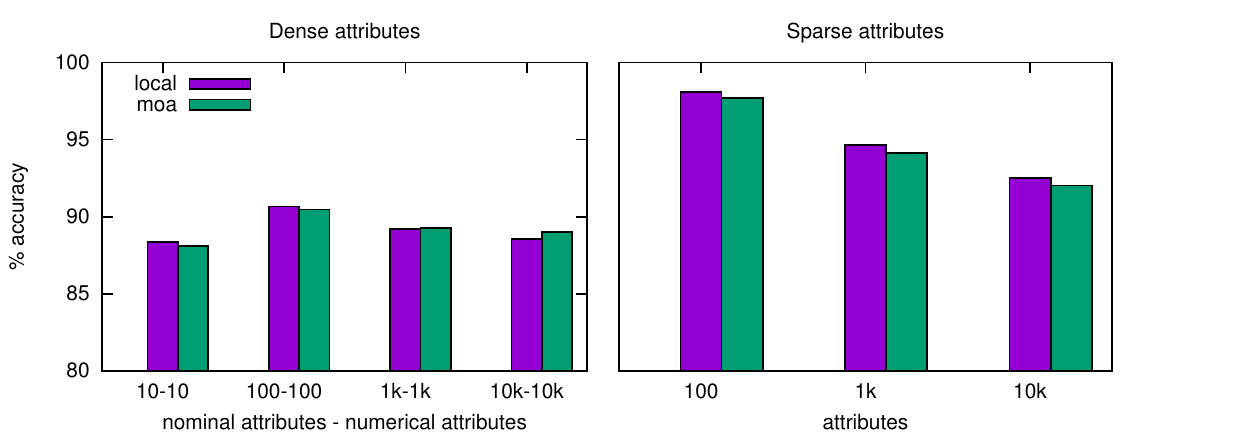}
\includegraphics[scale=1.05]{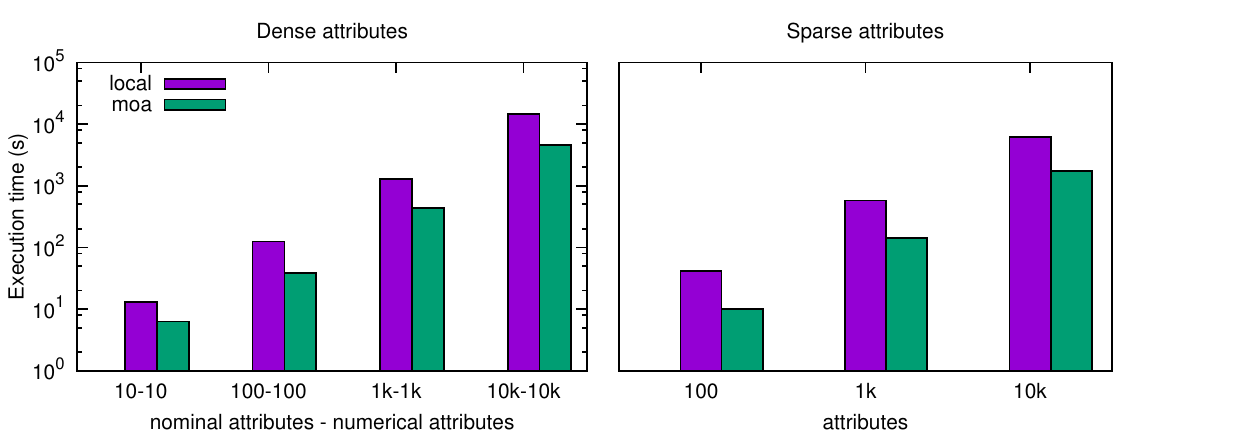}
\caption{Accuracy and execution time of \vht executed in local mode on \samoa compared to \moa, for dense and sparse datasets.}
\label{fig:dense_sparse_moa_vht_accuracy_time}
\end{figure}

In this first set of experiments, we test if \vht is performing as well as its counterpart hoeffding tree in \moa.
This is mostly a sanity check to confirm that the algorithm used to build the \vht does not affect the performance of the tree when all instances are processed sequentially by the model.
To verify this fact, we execute \vht local and \moa with both dense and sparse instances.
Figure~\ref{fig:dense_sparse_moa_vht_accuracy_time} shows that \vht local achieves the same accuracy as \moa, even besting it at times.
However, \vht local always takes longer than \moa to execute.
Indeed, the local execution engine of \samoa is optimized for simplicity rather than speed.
Therefore, the additional overhead required to interface \vht to DSPEs is not amortized by scaling the algorithm out.
Future optimized versions of \vht and the local execution engine should be able to close this gap.

\subsubsection{Accuracy of \vht local vs. distributed.}

Next, we compare the performance of \vht local with \vht built in a distributed fashion over multiple processors for scalability.
We use up to $p=8$ parallel statistics, due to memory restrictions, as our setup runs on a single machine.
In this set of experiments we compare the different versions of \vht, \wok and \wkz, to understand what is the impact of keeping instances for training after a model's split.
Accuracy of the model might be affected, compared to the local execution, due to delays in the feedback loop between statistics and model.
That is, instances arriving during a split are classified using an older version of the model compared to the sequential execution.
As our target is a distributed system where independent processes run without coordination, this delay is a characteristic of the algorithm as much as of the distributed SPE we employ.

\begin{figure}[t]
\centering
\includegraphics[scale=0.9]{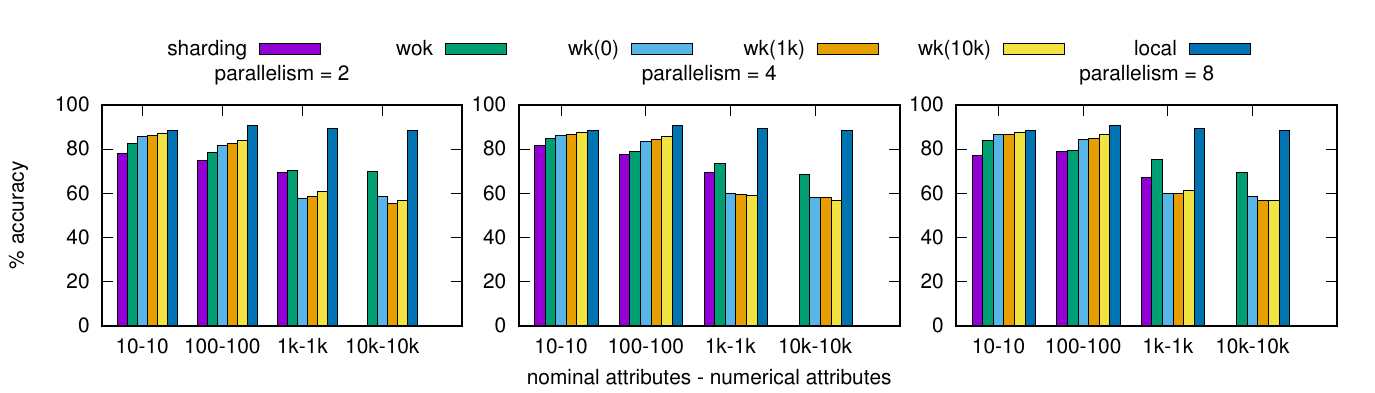}
\caption{Accuracy of several versions of \vht (local, \wok, \wkz) and sharding, for dense datasets.}
\label{fig:dense_accuracy_withlocal}
\end{figure}

We expect that buffering instances and replaying them when a split is decided would improve the accuracy of the model.
In fact, this is the case for dense instances with a small number of attributes (i.e., around $200$), as shown in Figure~\ref{fig:dense_accuracy_withlocal}.
However, when the number of available attributes increases significantly, the load imposed on the model seems to outweigh the benefits of keeping the instances for replaying.
We conjecture that the increased load in computing the splitting criterion in the statistics further delays the feedback to compute the split.
Therefore, a larger number of instances are classified with an older model, thus, negatively affecting the accuracy of the tree.
In this case, the additional load imposed by replaying the buffer further delays the split decision.
For this reason, the accuracy for \vht \wkz drops by about $30\%$ compared to \vht local.
Conversely, the accuracy of \vht \wok drops more gracefully, and is always within $18\%$ of the local version.

\begin{figure}[t]
\centering
\includegraphics[scale=1.05]{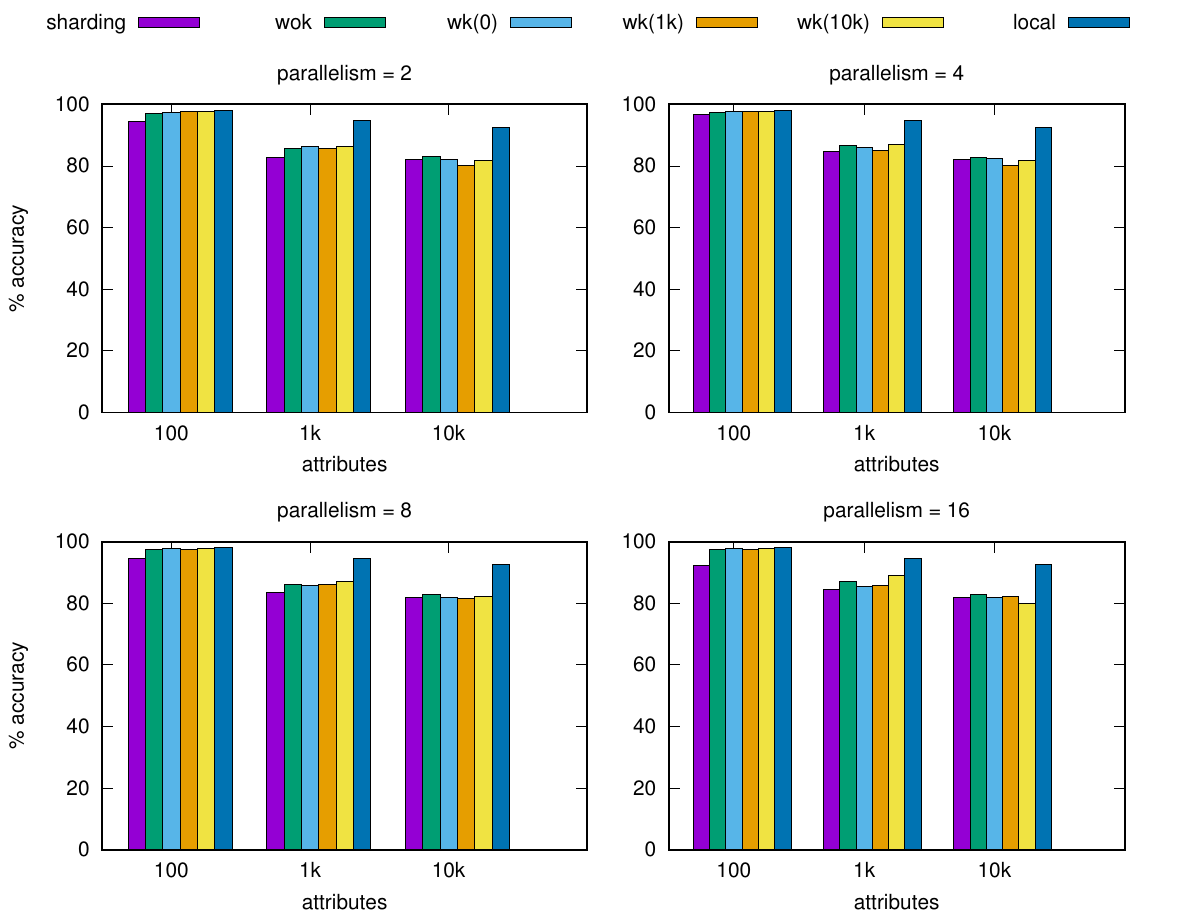}
\caption{Accuracy of several versions of \vht (local, \wok, \wkz) and sharding, for sparse datasets.}
\label{fig:sparse_accuracy_withlocal}
\end{figure}

\vht always performs approximatively $10\%$ better than sharding.
For dense instances with a large number of attributes (20k), sharding fails to complete due to its memory requirements exceeding the available memory.
Indeed, sharding builds a full model for each shard, on a subset of the stream.
Therefore, its memory requirements are $p$ times higher than a standard hoeffding tree.

When using sparse instances, the number of attributes per instance is constant, 
while the dimensionality of the attribute space increases.
In this scenario, increasing the number of attributes does not put additional load on the system.
Indeed, Figure~\ref{fig:sparse_accuracy_withlocal} shows that the accuracy of all versions is quite similar, and close to the local one.
This observation is in line with our conjecture that the overload on the system is the cause for the drop in accuracy on dense instances.

\begin{figure}[t]
\centering
\includegraphics[scale=1.1]{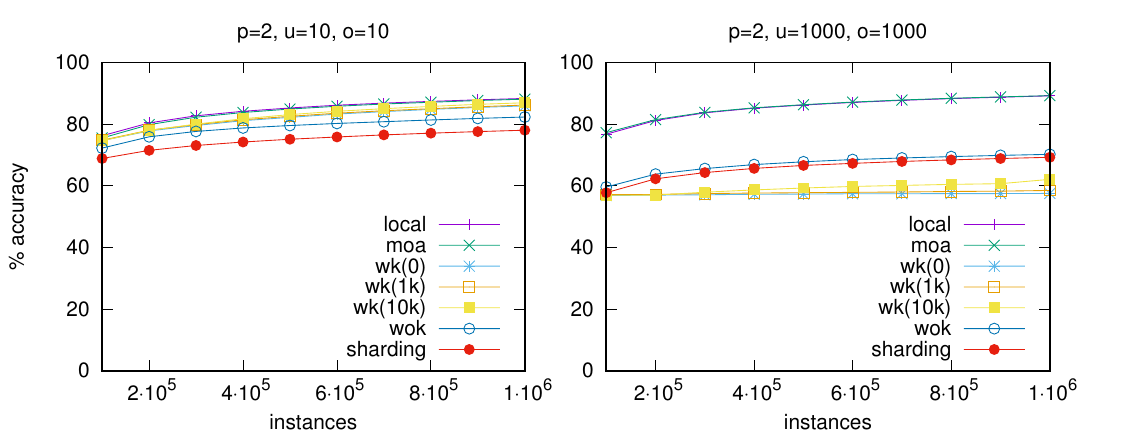}
\caption{Evolution of accuracy with respect to instances arriving, for several versions of \vht (local, \wok, \wkz) and sharding, for dense datasets.}
\label{fig:dense_accuracy_evolution}
\end{figure}

\begin{figure}[t]
\centering
\includegraphics[scale=1.1]{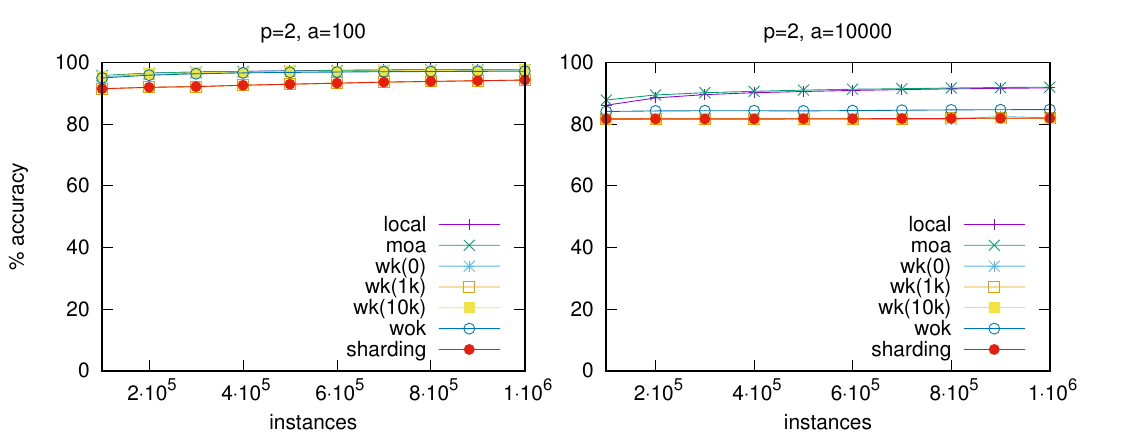}
\caption{Evolution of accuracy with respect to instances arriving, for several versions of \vht (local, \wok, \wkz) and sharding, for sparse datasets.}
\label{fig:sparse_accuracy_evolution}
\end{figure}

We also study how the accuracy evolves over time.
In general, the accuracy of all algorithms is rather stable, as shown in Figures~\ref{fig:dense_accuracy_evolution} and~\ref{fig:sparse_accuracy_evolution}.
For instances with 10 to 100 attributes, all algorithms perform similarly.
For dense instances, the versions of \vht with buffering, \wkz, outperform \wok, which in turn outperforms sharding.
This result confirms that buffering is beneficial for small number of attributes.
When the number of attributes increases to a few thousand per instance, the performance of these more elaborate algorithms drops considerably.
However, the \vht \wok continues to perform relatively well and better than sharding.
This performance, coupled with good speedup over \moa (as shown next) makes it a viable option for streams with a large number of attributes and a large number of instances. 

\begin{figure}
\centering
\includegraphics[scale=1.05]{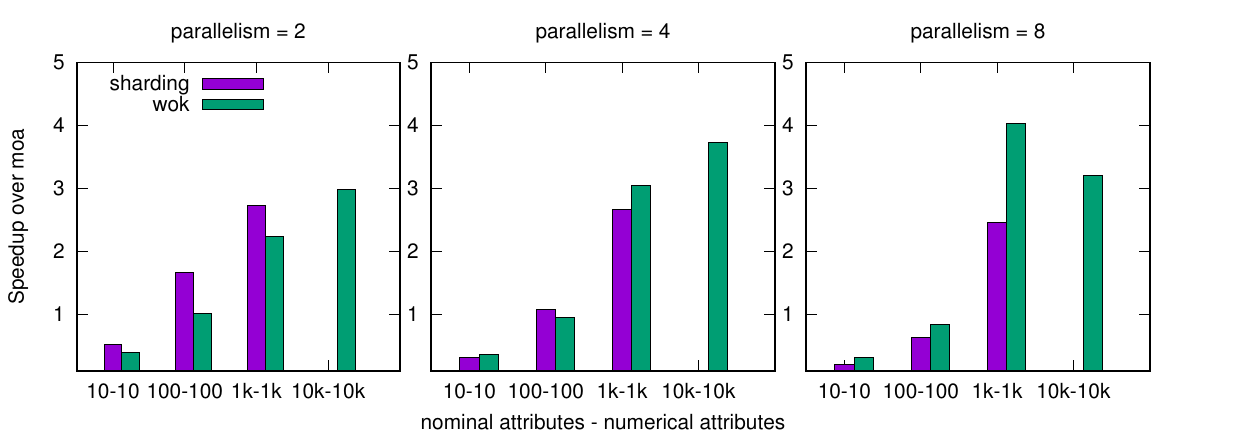}
\caption{Speedup of \vht \wok executed on \samoa compared to \moa for dense datasets.}
\label{fig:dense_speedup_withmoa}
\end{figure}

\spara{Speedup of \vht distributed vs. \moa.}
Since the accuracy of \vht \wkz is not satisfactory for both types of instances, next we focus our investigation on \vht \wok.
Figure~\ref{fig:dense_speedup_withmoa} shows the speedup of \vht for dense instances.
\vht \wok is about 2-10 times faster than \vht local and up to 4 times faster than \moa.
Clearly, the algorithm achieves a higher speedup when more attributes are present in each instance, as ($i$) there is more opportunity for parallelization, and ($ii$) the implicit load shedding caused by discarding instances during splits has a larger effect.
Even though sharding performs well in speedup with respect to \moa on small number of attributes, it fails to build a model for large number of attributes due to running out of memory.
In addition, even for a small number of attributes, \vht \wok outperforms sharding with a parallelism of $8$.
Thus, it is clear from the results that the vertical parallelism used by \vht offers better scaling behavior than the horizontal parallelism used by sharding.

\begin{figure}[t]
\centering
\includegraphics[scale=1.1]{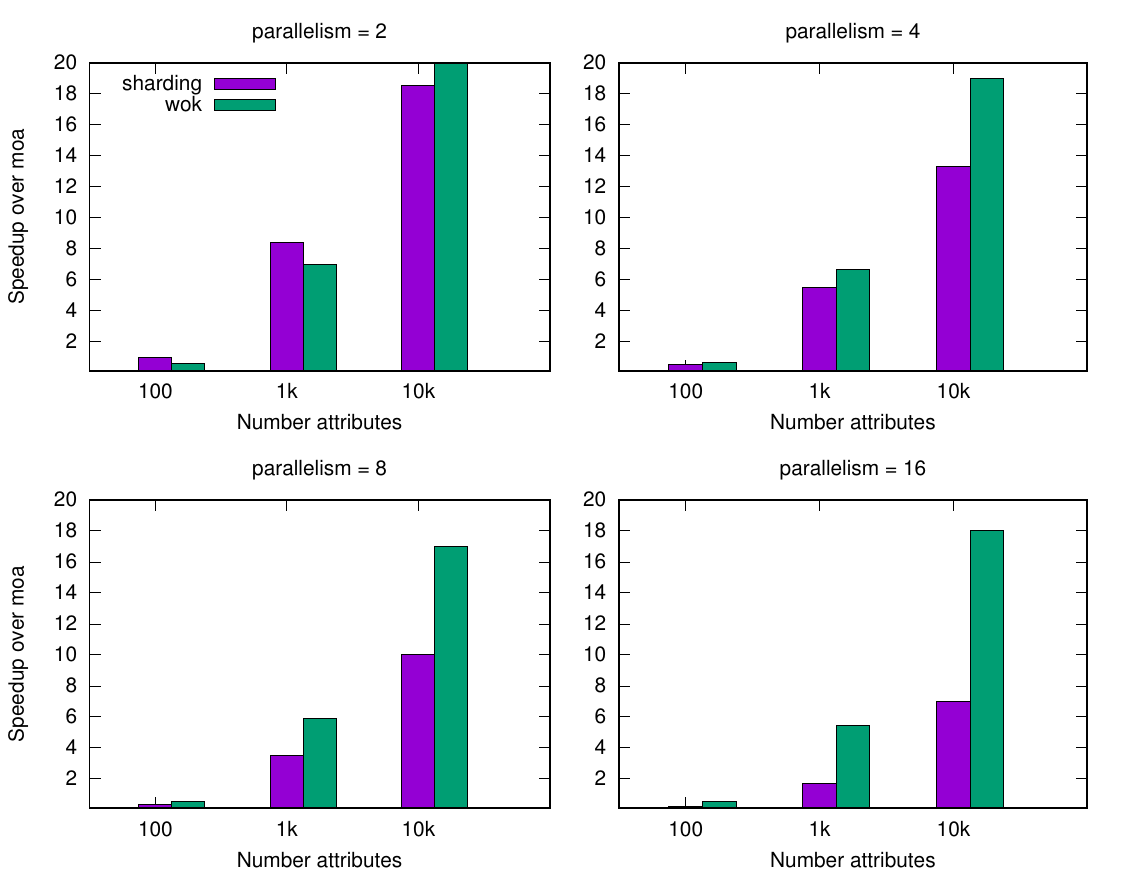}
\caption{Speedup of \vht \wok executed on \samoa compared to \moa for sparse datasets.}
\label{fig:sparse_speedup_withmoa}
\end{figure}

When testing the algorithms on sparse instances, as shown in Figure~\ref{fig:sparse_speedup_withmoa}, we notice that \vht \wok can reach up to 60 times the throughput of \vht local and 20 times the one of \moa (for clarity we only show the results with respect to \moa).
Similarly to what observed for dense instances, a higher speedup is observed when a larger number of attributes are present for the model to process.
This very large, superlinear speedup ($20 \times$ with $p=2$), is due to the aggressive load shedding implicit in the \wok version of \vht.
The algorithm actually performs consistently less work than the local version and \moa.

However, note that for sparse instances the algorithm processes a constant number of attributes, albeit from an increasingly larger space.
Therefore, in this setup, \wok has a constant overhead for processing each sparse instance, differently from the dense case.
\vht \wok outperforms sharding in most scenarios and especially for larger numbers of attributes and larger parallelism.

Increased parallelism does not impact accuracy of the model (see Figure~\ref{fig:dense_accuracy_withlocal} and Figure~\ref{fig:sparse_accuracy_withlocal}), but its throughput is improved.
Boosting the parallelism from 2 to 4 makes \vht \wok up to 2 times faster.
However, adding more processors does not improve speedup, and in some cases there is a slowdown due to additional communication overhead (for dense instances).
Particularly for sparse instances, parallelism does not impact accuracy which enables handling large sparse data streams while achieving high speedup over \moa.

\spara{Performance on real-world datasets.}
Tables~\ref{tab:real-accuracy} and~\ref{tab:real-time} show the performance of \vht, either running in a local mode or in a distributed fashion over a storm cluster of a few processors.
We also test two versions of \vht: \wok and wk(0).
In the same tables we compare \vht's performance with \moa and sharding.

The results from these real datasets demonstrate that \vht can perform similarly to \moa with respect to accuracy and at the same time process the instances faster.
In fact, for the larger dataset, covtypeNorm, \vht \wok exhibits 1.8 speedup with respect to \moa, even though the number of attributes is not very large (54 numeric attributes).
\vht \wok also performs better than sharding, even though the latter is faster in some cases.
However, the speedup offered by sharding decreases when the parallelism level is increased from 2 to 4 shards.

\begin{table}[htbp]
\vspace{-5mm}
\caption{Average accuracy (\%) for different algorithms, with parallelism level (p), on the real-world datasets.}
\vspace{-\baselineskip}
\begin{center}
\small
\tabcolsep=0.5em
\begin{tabular}{l r r r r r r r r}
\toprule
dataset	&	\moa	&	\multicolumn{5}{c}{\vht}	&	\multicolumn{2}{c}{Sharding}		\\
		&		&	local	&	\wok	&	\wok	&	wk(0)&	wk(0)&		&		\\
		&		&		&	p=2	&	p=4	&	p=2	&	p=4	&	p=2	&	p=4	\\
\midrule
elec		&	75.4	&	75.4	&	75.0	&	75.2	&	75.4	&	75.6	&	74.7	&	74.3	\\
phy		&	63.3	&	63.8	&	62.6	&	62.7	&	63.8	&	63.7	&	62.4	&	61.4	\\
covtype	&	67.9	&	68.4	&	68.0	&	68.8	&	67.5	&	68.0	&	67.9	&	60.0	\\
\bottomrule
\end{tabular}
\end{center}
\label{tab:real-accuracy}
\end{table}%

\begin{table}[htbp]
\vspace{-5mm}
\caption{Average execution time (seconds) for different algorithms, with parallelism level (p), on the real-world datasets.}
\vspace{-\baselineskip}
\begin{center}
\small
\tabcolsep=0.5em
\begin{tabular}{l r r r r r r r r}
\toprule
Dataset	&	\moa		&	\multicolumn{5}{c}{\vht}	&	\multicolumn{2}{c}{Sharding}									\\
		&			&	local		&	\wok		&	\wok		&	wk(0)	&	wk(0)	&			&			\\
		&			&			&	p=2		&	p=4		&	p=2		&	p=4		&	p=2		&	p=4		\\
\midrule
elec		&	1.09		&	1		&	2		&	2		&	2		&	2		&	2		&	2.33		\\
phy		&	5.41		&	4		&	3.25		&	4		&	3		&	3.75		&	3		&	4		\\
covtype	&	21.77	&	16		&	12		&	12		&	13		&	12		&	9		&	11		\\
\bottomrule
\end{tabular}
\end{center}
\label{tab:real-time}
\vspace{-10mm}
\end{table}%

\subsection{Summary}

In conclusion, our \vht algorithm has the following performance traits.
We learned that for a small number of attributes, it helps to buffer incoming instances that can be used in future decisions of split.
For larger number of attributes, the load in the model can be high and larger delays can be observed in the integration of the feedback from the local statistics into the model.
In this case, buffered instances may not be used on the most up-to-date model and this can penalize the overall accuracy of the model.

With respect to a centralized sequential tree model (\moa), it processes dense instances with thousands of attributes up to $4\times$ faster with only $10-20\%$ drop in accuracy.
It can also process sparse instances with thousands of attributes up to $20\times$ faster with only $5-10\%$ drop in accuracy.
Also, its ability to build the tree in a distributed fashion using tens of processors allows it to scale and accommodate thousands of attributes and parse millions of instances.
Competing methods cannot handle these data sizes due to increased memory and computational complexity.

\section{Distributed AMRules}
\label{sec:amrules}

Decision rule learning is a category of machine learning algorithms whose goal is to extract a set of decision rules from the training data.
These rules are later used to predict the unknown \emph{label} values for test data.
A rule is a logic expression of the form:
$$\textbf{IF} \: antecedent \: \textbf{THEN} \: consequent$$
or, equivalently, $head \leftarrow body$, where \emph{head} and \emph{body} correspond to the \emph{consequent} and \emph{antecedent}, respectively.

The body of a rule is a conjunction of multiple clauses called \emph{features}, each of which is a condition on an attribute of the instances. Such conditions consist of the identity of an attribute, a threshold value and an operator. 
For instance, the feature ``$x < 5$'' is a condition on attribute \emph{x}, with threshold value \emph{5} and operator \emph{less-than} ($<$).
An instance is said to be \emph{covered} by a rule if its attribute values satisfy all the features in the rule body.
The head of the rule is a function to be applied on the covered instances to determine their label values.
This function can be a constant or a function of the attributes of the instances, e.g., 
$ax + b \leftarrow x < 5 $.

AMRules is an algorithm for learning regression rules on streaming data.
It incrementally constructs the rule model from the incoming data stream.
The rule model consists of a set of \emph{normal} rules (which is empty at the beginning), and a \emph{default} rule.
Each normal rule is composed of 3 parts: a \emph{body} which is a list of features, a \emph{head} with information to compute the prediction for those instance covered by the rule, and \emph{statistics} of past instances to decide when and how to add a new feature to its body.
In fact, the default rule is a rule with an empty \emph{body}.

For each incoming instance, AMRules searches the current rule set for those rules that cover the instance.
If an instance is not covered by any rule in the set, it is considered as being covered by the default rule.
The heads of the rules are first used to compute the prediction for the instance they cover.
Later, their statistics are updated with the attribute values and label value of the instance.
There are two possible modes of operation: ordered and unordered.
In ordered-rules mode, the rules are checked according to the order of their creation, and only the first rule is used for prediction and then updated.
In unordered-rules mode, all covering rules are used and updated.
In this work, we focus on the former which is more often used albeit more challenging to parallelize.

Each rule tries to expand its body after it receives $N_{m}$ updates.
In order to decide on the feature to expand, the rule incrementally computes a standard deviation reduction (SDR) measure~\cite{fimt-sdr} for each potential feature.
Then, it computes the $ratio$ of the second-largest SDR value over the largest SDR value.
This $ratio$ is used with a high confidence interval $\epsilon$ computed using the Hoeffding bound~\cite{Hoeffding1963bound} to decide to expand the rule or not:
if $ratio + \epsilon < 1$, the rule is expanded with the feature corresponding to the largest SDR value.
Besides, to avoid missing a good feature when there are two (or more) equally good ones, rules are also expanded if the Hoeffding bound $\epsilon$ falls below a threshold.
If the default rule is expanded, it becomes a \emph{normal} rule and is added to the rule set.
A new default rule is initialized to replace the previous one.

Each rule records its prediction error and applies a modified version of the Page-Hinkley test~\cite{pagehinkley} for streaming data to detect changes.
If the test indicates that the cumulative error has exceeded a threshold, the rule is evicted from the rule set.
The algorithm also employ outlier detection to check if an instance, although being covered by a rule, is an anomaly.
If an instance is deemed as an anomaly, it is treated as if the rule does not cover it and is checked against other rules.
The following sections describe two possible strategies to parallelize AMRules that are implemented in \samoa.

\begin{figure}[t]
\begin{center}
\includegraphics[width=0.51\columnwidth]{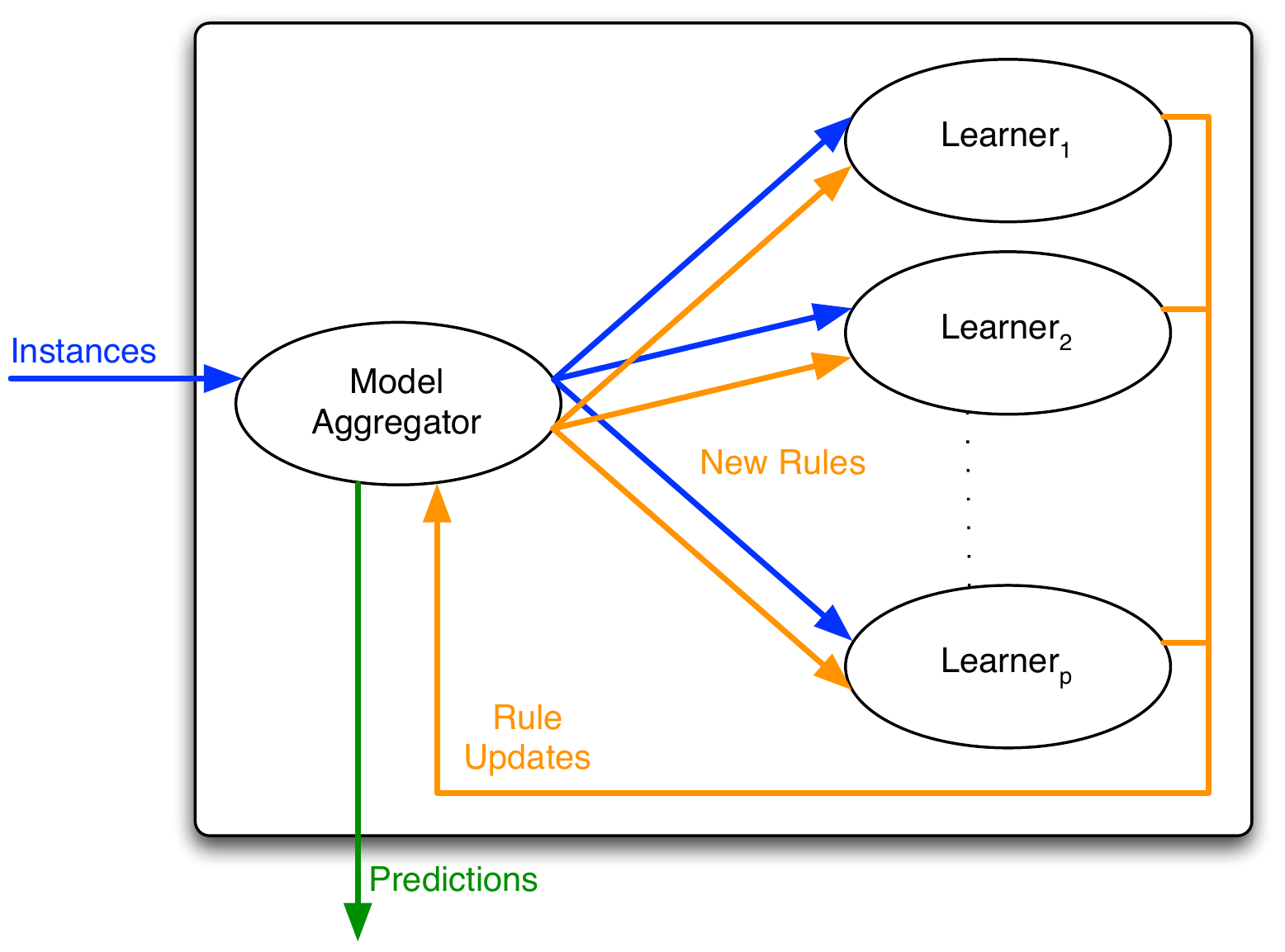}
\hspace{-2mm}
\includegraphics[width=0.49\columnwidth]{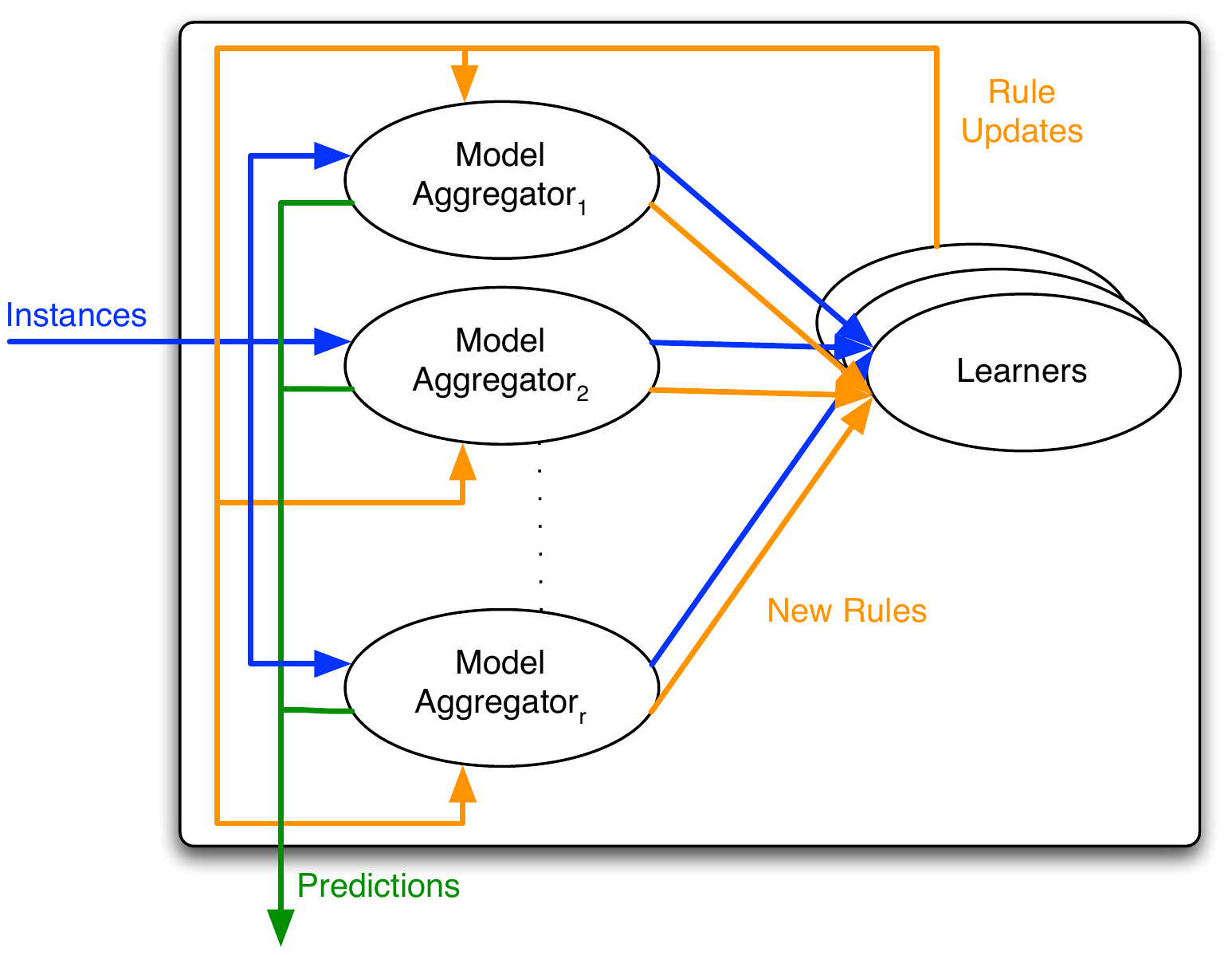}
\caption{(Left) Vertical AMRules (VAMR). (Right) AMRules with multiple horizontally parallelized Model Aggregators.}
\label{fig:amrules-vertical_horizontal1}
\end{center}
\end{figure}

\subsection{Vertical Parallelism}
In AMRules, each rule can evolve independently, as its expansion is based solely on the statistics of instances it covers.
Also, searching for the best feature among all possible ones in an attempt to expand a rule is computationally expensive.

Given these observations, we decide to parallelize AMRules by delegating the training process of rules to multiple \emph{learner processors}, each of which handles only a subset of the rules. 
Besides the learners, a \emph{model aggregator processor} is also required to filter and redirect the incoming instances to the correct learners.
The aggregator manages a set of simplified rules that have only head and body, i.e., do not keep statistics.
The bodies are used to identify the rules that cover an instance, while the heads are used to compute the prediction.
Each instance is forwarded to the designated learners by using the ID of the covering rule.
At the learners, the corresponding rules' statistics are updated with the forwarded instance.
This parallelization scheme guarantees that the rules created are the same as in the sequential algorithm.
Figure~\ref{fig:amrules-vertical_horizontal1}(left) depicts the design of this vertically parallelized version of AMRules, or Vertical AMRules (VAMR for brevity).

The model aggregator also manages the statistics of the default rule, and updates it with instances not being covered by any other rule 
in the set.
When the default rule is expanded and adds a new rule to the set, the model aggregator sends a message with the newly added rule to one of the learners, which is responsible for its management.
The assignment of a rule to a learner is done based on the rule's ID via key grouping.
All subsequent instances that are covered by this rule are forwarded to the same learner. 

At the same time, learners update the statistics of each corresponding rule with each processed instance.
When enough statistics have been accumulated and a rule is expanded, the new feature is sent to the model aggregator to update the body of the rule. 
Learners can also detect changes and remove existing rules.
In such an event, learners inform the model aggregator with a message containing the removed rule ID.

As each rule is replicated in the model aggregator and in one of the learners, their bodies in model aggregator might not be up-to-date.
The delay between rule expansion in the learner and model update in the aggregator depends mainly on the queue length at the model aggregator.
The queue length, in turn, is proportional to the volume and speed of the incoming data stream.
Therefore, instances that are in the queue before the model update event might be forwarded to a recently expanded rule which no longer covers the instance.

Coverage test is performed again at the learner, thus the instance is dropped if it was incorrectly forwarded.
Given this additional test, and given that rule expansion can only increase the selectivity of a rule, when using unordered rules the accuracy of the algorithm is unaltered.
However, in ordered-rules mode, these temporary inconsistencies might affect the statistics of other rules because the instance should have been forwarded to a different rule.

\subsection{Horizontal Parallelism}
A bottleneck in VAMR is the centralized model aggregator.
Given that there is no straightforward way to vertically parallelize the execution of the model aggregator while maintaining the order of the rules, we explore an alternative based on horizontal parallelism.
Specifically, we introduce multiple replicas of the model aggregator, so that each replica maintains the same copy of the rule set but processes only a portion of the incoming instances.

\spara{Horizontally parallelized model aggregator.}
The design of this scheme is illustrated in Figure~\ref{fig:amrules-vertical_horizontal1}(right).
The basic idea is to extend VAMR and accommodate multiple model aggregators into the design.
Each model aggregator still has a rule set and a default rule.
The behavior of this scheme is similar to VAMR, except that each model aggregator now processes only a portion of the input data, i.e., the amount of instances each of them receives is inversely proportional to the number of model aggregators.
This affects the prediction statistics and, most importantly, the training statistics of the default rules. 

Since each model aggregator processes only a portion of the input stream, each default rule is trained independently with different 
portions of the stream.
Thus, these default rules evolve independently and potentially create overlapping or conflicting rules. 
This fact also introduces the need for a scheme to synchronize and order the rules created by different model aggregators.
Additionally, at the beginning, the scheme is less reactive compared to VAMR as it requires more instances for the default rules to start 
expanding.
Besides, as the prediction function of each rule is adaptively constructed based on attribute values and label values of past instances, having only a portion of the data stream leads to having less information and potentially lower accuracy.
We show how to address these issues next.

\begin{figure}[t]
\begin{center}
\includegraphics[width=0.49\columnwidth]{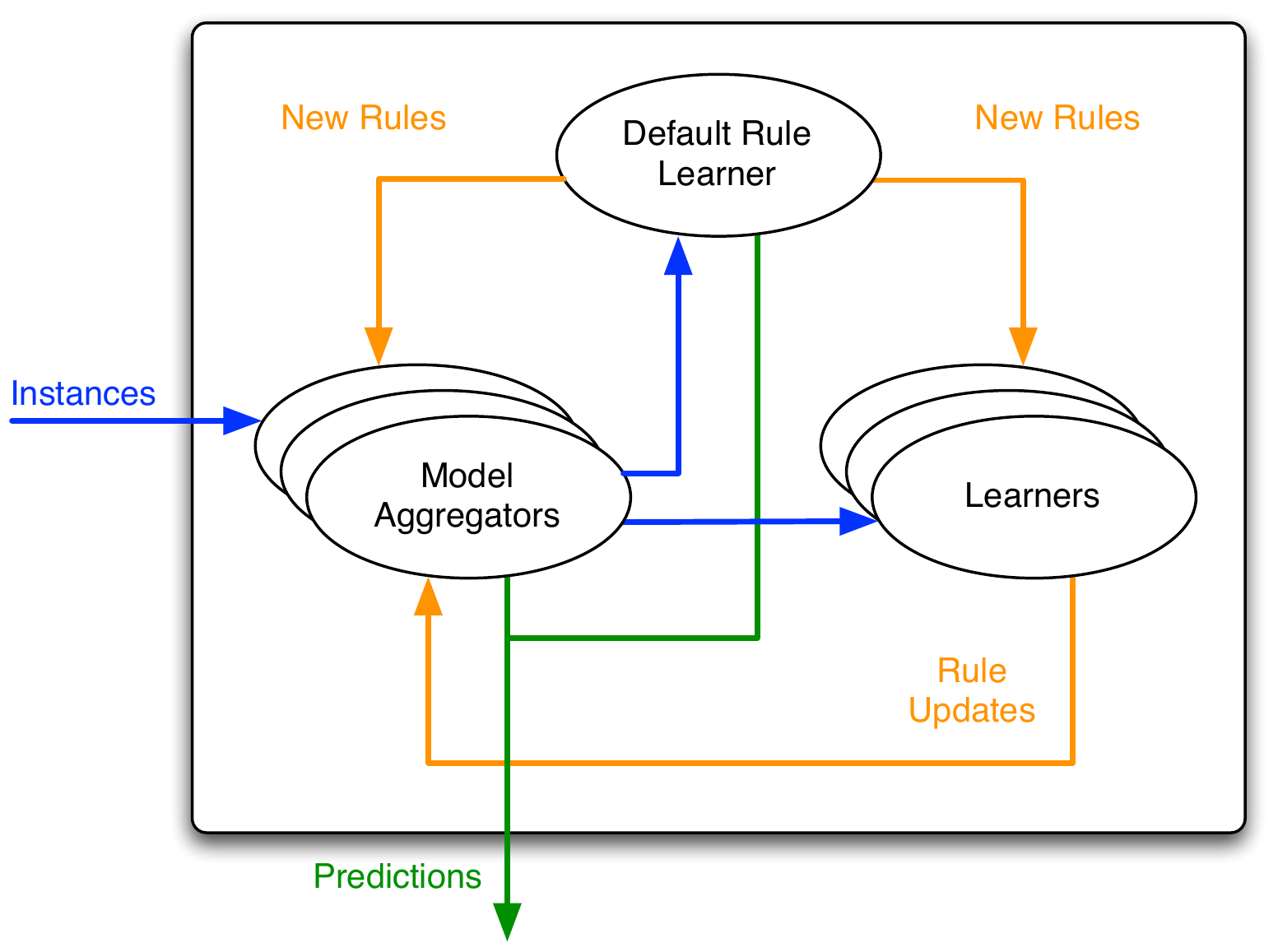}
\includegraphics[width=0.49\columnwidth]{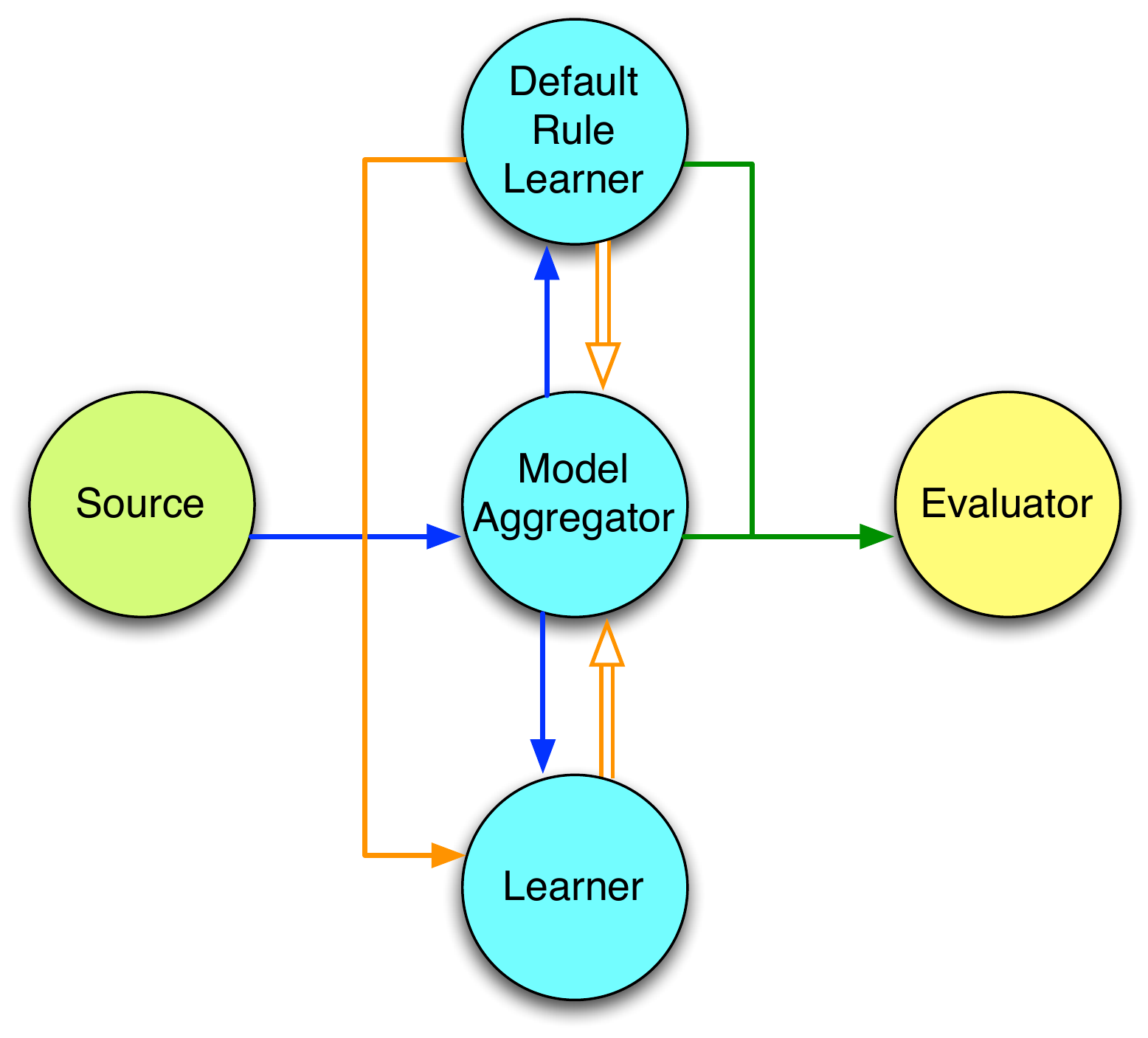}
\caption{
(Left) Hybrid AMRules (HAMR) with multiple Model Aggregators and separate Default Rule Learner.
(Right) Prequential evaluation task for HAMR. Single lines represent single messages (\emph{key grouping}, \emph{shuffle grouping}) while double lines represent broadcast messages (\emph{all grouping}).}
\label{fig:amrules-horizontal2-hamr-task}
\end{center}
\vspace{-4mm}
\end{figure}

\spara{Centralized rule creation.}
In order to address the issues with distributed creation of rules, we move the default rule in model aggregators to a specialized default rule learner processor.
With the introduction of this new component, some modifications are required in the model aggregators, but the behavior of the learners is still the same as in VAMR.
However, as a result, all the model aggregators are in synch.

As the default rule is now moved to the designated learner, those instances that are not covered by any rules are forwarded from the model aggregators to this learner.
This specialized learner updates its statistics with the received instances and, when the default rule expands, it broadcasts the newly created rule to the model aggregators.
The new rule is also sent to the assigned learner, as determined by the rule's ID. 

The components of this scheme are shown in Figure~\ref{fig:amrules-horizontal2-hamr-task}(left), where this scheme is referred as Hybrid AMRules (HAMR), as it is a combination of vertical and horizontal parallelism strategies.

\subsection{Evaluation}
\label{sec:eval}

We evaluate the performance of the 2 distributed implementations of AMRules, i.e., VAMR and HAMR, in comparison to the centralized implementation in MOA\footnote{\url{http://moa.cms.waikato.ac.nz}} (MAMR).

\spara{Evaluation methodology.}
We plug VAMR and HAMR into a \emph{prequential evaluation task}~\cite{GamaSR13},
where each instance is first used to test and then to train the model.
This evaluation task includes a \emph{source} processor which provides the input stream and a \emph{evaluator} processor which records the rate and accuracy of prediction results.
The final task for HAMR is depicted in Figure~\ref{fig:amrules-horizontal2-hamr-task}(right).
The parallelism level of the model is controlled by setting the number of learners $p$ and the number of model aggregators $r$.
The task for VAMR is similar but the default rule learner is excluded and model aggregator's parallelism level is always 1.
Each task is repeated for five runs. 

\spara{Datasets.}
We perform the same set of experiments with 3 different datasets, i.e., \emph{electricity}, \emph{airlines} and \emph{waveform}.
\begin{squishlist}
\item \emph{Electricity}: A dataset from the UCI Machine Learning Repository~\cite{ucidata} which records the electricity power consumption (in watt-hour) of a household from December 2006 to November 2010. The dataset contains more than 2 millions 12-attribute records.
\item \emph{Airlines}: A dataset recording the arrival delay (in seconds) of commercial flights within the USA in year 2008\footnote{\url{http://kt.ijs.si/elena_ikonomovska/data.html}}. 
It contains more than 5.8 millions records, each with 10 numerical attributes.
\item \emph{Waveform}: A dataset generated using an artificial random generator.
To generate an instance, it picks a random waveform among the 3 available ones, and 21 attribute values for this instance are generated according to the chosen waveform.
Another 19 noise attributes are generated and included in the new instance, making a total of 40 attributes for each instance.
The label value is the index of the waveform (0, 1, or 2).
Although this dataset does not fit perfectly to the definition of a regression task, it allows us to test our implementations with a high number of numerical attributes.
\end{squishlist}

\spara{Setup.}
The evaluation is performed on an OpenStack\footnote{\url{http://www.openstack.org}} cluster of 9 nodes, each with 2 Virtual CPUs @ 2.3GHz and 4GB of RAM.
All the nodes run Red Hat Enterprise Linux 6.
The distributed implementations are evaluated on a Samza\footnote{\url{http://samza.incubator.apache.org}} cluster with Hadoop YARN 2.2\footnote{\url{http://hadoop.apache.org}} and Kafka 0.8.1.\footnote{\url{http://kafka.apache.org}}
The Kafka brokers coordinate with each other via a single-node ZooKeeper 3.4.3\footnote{\url{http://zookeeper.apache.org}} instance.
The replication factor of Kafka's streams in these experiments is 1.
The performance of MAMR is measured on one of the nodes in the cluster.

\begin{figure}[t]
\centering
\includegraphics{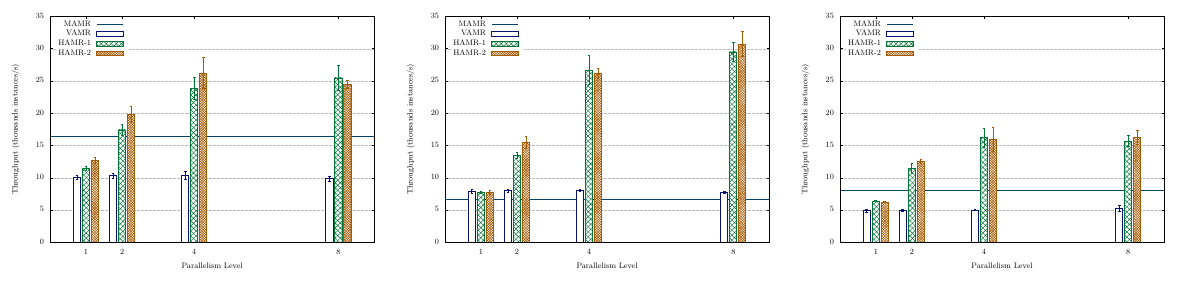}
 \caption{Throughput of distributed AMRules with \emph{electricity} (left), \emph{airlines} (mid), and \emph{waveform} (right).}%
 \label{fig:amr-throughput}%
\end{figure}

\spara{Throughput.}
\label{subsec:throughput}
The throughput of several variants of AMRules is shown in Figure~\ref{fig:amr-throughput} for each dataset examined.
HAMR-1 and HAMR-2 stand for \emph{HAMR with 1 learner} and \emph{HAMR with 2 learners}.
The parallelism levels of VAMR represents the number of its learners, while in the case of HAMR it represents the number of model aggregators.

With \emph{electricity} and \emph{waveform}, the communication overhead in VAMR (to send instances from model 
aggregator to learners and evaluator) exceeds the throughput gained from delegating the training process to the learners
and results in a lower overall throughput compared to MAMR's.
However, with \emph{airlines}, the performance of VAMR is better than MAMR.
To verify that the training process for \emph{airlines} is more computationally intensive than the other two datasets and, thus, VAMR is more effective for this dataset, we compare the statistics of rules and predicates creation for the 3 datasets.
The number of rules created, rules removed, and features created (when a rule is expanded) by MAMR with the 3 datasets are presented in Table~\ref{tab:rules-statistics}.
By subtracting the total number of rules removed from the total number of rules created, we can have an estimation of the average number of rules in the model for each dataset.
A higher number of rules in the model and a higher number of features created suggest that the model is more complex and it takes more processing power to search for the best new feature when a rule attempts to expand. 

Although VAMR can perform better than MAMR for computationally intensive datasets, its throughput does not change with different parallelism level.
This is due to the bottleneck at the model aggregator.
The processing of each instance at the model aggregator consists of three steps: finding the covering rules, forwarding the instance to the corresponding learners, and applying covering rules to predict the label value of the instance.
Since the complexity of these three steps for an instance is constant, the final throughput is unaffected by the parallelism level.

\begin{table}[t]
\centering
\caption{Statistics for features and rules (with MAMR) for different datasets.}
\tabcolsep=2mm
\begin{tabular}{l r r r}
\toprule
&	Electricity	&	Airlines	&	Waveform \\
\midrule
Instances				&	\num{2049280}	&	\num{5810462}	&	\num{1000000}	\\
\# Attributes			&	12	&	10	&	40	\\
Result message size (B)	&	891	&	764	&	1446 	\\
\# Rules created		&	1203	&	2501	&	270	\\
\# Rules removed		&	1103	&	1040	&	51	\\
Avg. \# Rules	&	100	&	1461	&	219	\\
\# Features created		&	 \num{1069}	&	\num{10606}	&	\num{1245}	\\
\bottomrule
\end{tabular}
\label{tab:rules-statistics}
\end{table}

\begin{figure}[t]
\centering
\includegraphics{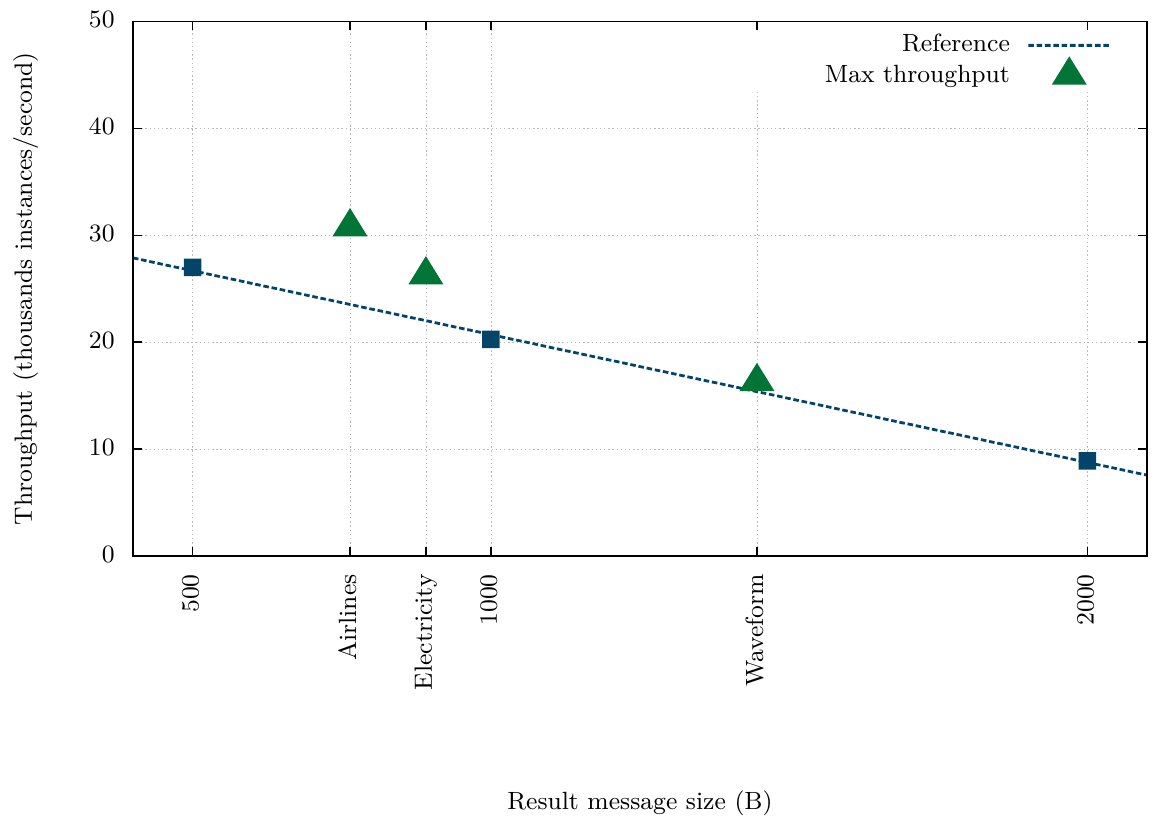}
 \caption{Maximum throughput of HAMR vs message size.}%
 \label{fig:max-throughput}%
\end{figure}

\begin{figure*}[!ht]
        \centering
\includegraphics{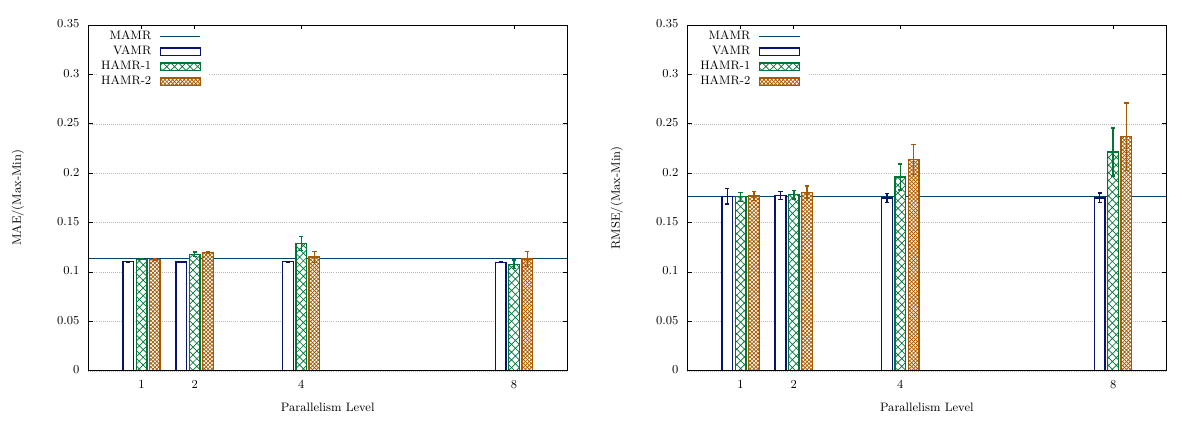}
        \caption{MAE (left) and RMSE (right) of distributed AMRules with the \emph{electricity} dataset.}%
\label{fig:electricity-error}%
\end{figure*}

The throughput of HAMR-1 and HAMR-2 exhibit a better scalability compared to VAMR. Up to parallelism level of 4, the throughput increases almost linearly with the number of model aggregators.
However, there is no or little improvement when this number is increased from 4 to 8.
As we measure this throughput at a single evaluator, we suspect that the bottleneck is in the maximum rate the evaluator can read from the output streams of the model aggregators and default rule learner.
To investigate this issue, we plot the maximum throughput of HAMR against the size of messages from model aggregators and default rule learner to evaluator in Figure~\ref{fig:max-throughput}.
The values of throughput of a single-partition Samza stream with messages of size 500B, 1000B and 2000B are used to compute the linear regression line (reference line) in the figure. 
The message size for different datasets is shown in Table~\ref{tab:rules-statistics}.

As reading is expected to be faster than writing in Samza and Kafka, the maximum rate the evaluator in HAMR can read from multiple 
stream partitions is expected to be higher than the throughput represented by the reference line.
This fact is reflected in Figure~\ref{fig:max-throughput} as the maximum throughput of HAMR for the 3 datasets constantly exceeds the reference line.
However, the difference between them is relatively small.
This result is a strong indicator that the bottleneck is the maximum reading rate of the evaluator.
If there is no need to aggregate the result from different streams, this bottleneck can be eliminated. 

\begin{figure*}[t]
        \centering
	\includegraphics{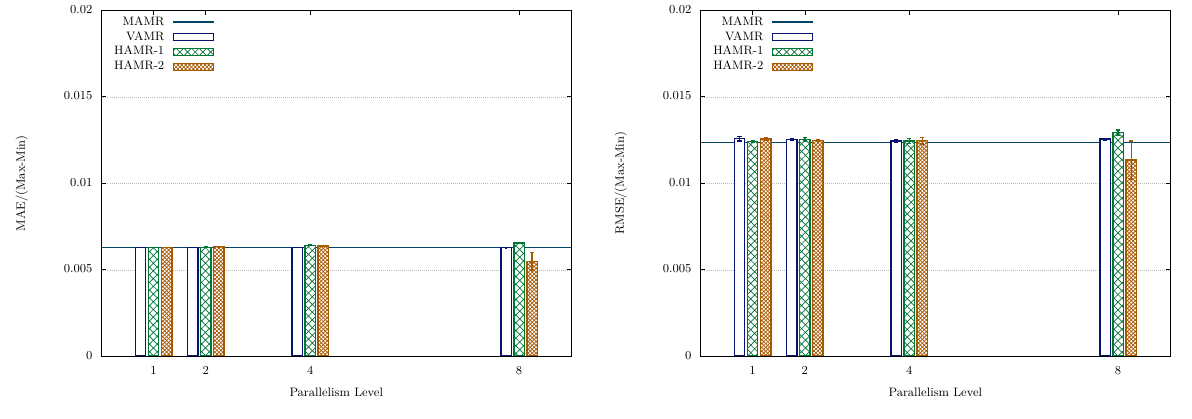}
	\caption{MAE (left) and RMSE (right) of distributed AMRules with the \emph{airlines} dataset.}%
\label{fig:airlines-error}%
\end{figure*}
 
\begin{figure*}[t]
        \centering
	\includegraphics{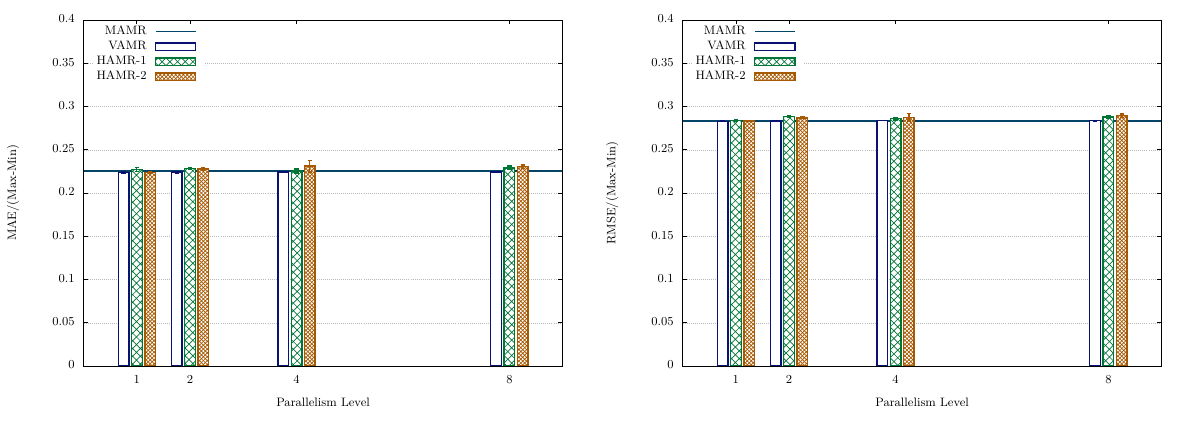}
	\caption{MAE (left) and RMSE (right) of distributed AMRules with the \emph{waveform} random generator.}%
\label{fig:waveform-error}
\end{figure*}

\spara{Accuracy.}
We evaluate accuracy of the different implementations of AMRules in terms of \emph{Mean Absolute Error} (MAE) and \emph{Root Mean Square Error} (RMSE).
Figure~\ref{fig:electricity-error}, \ref{fig:airlines-error} and \ref{fig:waveform-error} show the MAE and RMSE for the three datasets, normalized by the range of label values in each dataset.

Most of the figures show that error values of distributed implementations presents very small fluctuations around the corresponding MAMR error line. However, there is a significant increment in the value of RMSE in HAMR with \emph{electricity} dataset when 
the number of model aggregators is increased to 4 or 8. Moreover, larger variances, i.e., \emph{standard error} is greater than 
5\% of the average MAE or RMSE, are also observed in the case of higher parallelism levels ($p \ge 4$) of HAMR. 
The probable cause is that when a rule in the model aggregators is out-of-sync with the corresponding one in the learners, i.e.,
model aggregators are not yet updated with a newly created rule, the number of instances that use an outdated model is multiplied by the throughput. 

\begin{table}[ht]
\caption{Memory consumption of MAMR for different datasets.}
\centering
\begin{tabular}{l r r}
\toprule
Dataset & \multicolumn{2}{c}{Memory consumption (MB)} \\
\cmidrule(lr){2-3}
&	Avg.	&	Std. Dev. \\
\midrule
Electricity	& 52.4	& 2.1 \\
Airlines	& 120.7	& 51.1 \\
Waveform	& 223.5	& 8 \\
\bottomrule
\end{tabular}
\label{tab:mamr-memory}
\end{table}

\begin{table}[!ht]
\caption{Memory consumption of VAMR for different datasets and parallelism levels (AVG: average; SD: standard deviation).}
\begin{center}
\begin{tabular}{l c c c c c}
\toprule
Dataset	&	Parallelism	&	\multicolumn{4}{c}{Memory Consumption (MB)}	\\
							\cmidrule(lr){3-6}
		&				&	\multicolumn{2}{c}{Model Aggregator}	&	\multicolumn{2}{c}{Model Learner}	\\
							\cmidrule(lr){3-4} \cmidrule(lr){5-6}
		&		&	AVG	&	SD	&	AVG	& SD		\\
\midrule
Electricity \\
& 1 & 266.5 & 5.6 & 40.1 & 4.3 \\
& 2 & 264.9 & 2.8 & 23.8 & 3.9 \\
& 4 & 267.4 & 6.6 & 20.1 & 3.2\\
& 8 & 273.5 & 3.9 & 34.7 & 2.9\\
Airlines \\
& 1 & 337.6 & 2.8 & 83.6 & 4.1 \\
& 2 & 338.1 & 1.0 & 38.7 & 1.8 \\
& 4 & 337.3 & 1.0 & 38.8 & 7.1 \\
& 8 & 336.4 & 0.8 & 31.7 & 0.2 \\
Waveform \\
& 1 & 286.3 & 5.0 & 171.7 & 2.5 \\
& 2 & 286.8 & 4.3 & 119.5 & 10.4 \\
& 4 & 289.1 & 5.9 & 46.5 & 12.1 \\
& 8 & 287.3 & 3.1 & 33.8 & 5.7	\\
\bottomrule
\end{tabular}
\end{center}
\label{tab:vamr-memory}
\end{table}%

\spara{Memory.}
The reference memory consumption of MAMR for different datasets is presented in Table~\ref{tab:mamr-memory}. This table shows 
that \emph{waveform} consumes the largest amount of memory, followed by \emph{airlines} and then \emph{electricity}. 

Table~\ref{tab:vamr-memory} reports the memory consumption (\emph{average} and \emph{standard deviation}) of model aggregator and 
learner processors of VAMR for
the 3 datasets with different parallelism levels. First of all, we notice a high memory consumption at the model aggregator. However
this amount does not differ much for different datasets. This suggests that, at the model aggregator, there is a constant overhead in 
memory usage, but the memory consumption due to the growing decision model is small. 

Second, the memory consumption per learner decreases as more
learners are used. As there is definitely some memory overhead for each learner instance, there is no significant reduction of memory
usage per learner when the parallelism level goes from 4 to 8. However, the result indicates that we can spread a large decision model 
over multiple learners and make it possible to learn very large models whose size exceeds the memory capacity of a single 
machine.

\section{Conclusions}

We presented the \samoa platform for mining big data streams.
The platform supports the most common machine learning tasks such as classification, clustering, and regression.
It also provides a simple API for developers that allows to implement distributed streaming algorithms easily.
\samoa is already available and can be found online at:
\begin{center}
\url{https://samoa.incubator.apache.org}
\end{center}

The website includes a wiki, an API reference, and a developer's manual.
Several examples of how the software can be used are also available.
The code is hosted on GitHub.
Finally, \samoa is released as open source software under the Apache Software License v2.0.

\bibliographystyle{abbrvnat}
\bibliography{biblio,referencesv,references}

\end{document}